\documentclass[showpacs,preprintnumbers,amsmath,amssymb]{revtex4}
\usepackage{amsmath,amssymb,graphics,epsfig,subfigure}
\usepackage{color}

\begin{document}
\renewcommand{\baselinestretch}{1.3}

\title{Topology of equatorial timelike circular orbits around stationary black holes}

\author{Shao-Wen Wei$^{a,b}$ \footnote{weishw@lzu.edu.cn},
Yu-Xiao Liu$^{a,b}$ \footnote{liuyx@lzu.edu.cn}}

\affiliation{$^{a}$Lanzhou Center for Theoretical Physics, Key Laboratory of Theoretical Physics of Gansu Province, School of Physical Science and Technology, Lanzhou University, Lanzhou 730000, People's Republic of China,\\
$^{b}$Institute of Theoretical Physics $\&$ Research Center of Gravitation,
Lanzhou University, Lanzhou 730000, People's Republic of China}

\begin{abstract}
A topological approach has been successfully used to study the properties of the light ring and the null circular orbit, in a generic black hole background. However, for the equatorial timelike circular orbit, quite different from the light ring case, its radius is closely dependent of the energy and angular momentum of a test particle. This fact seems to restrict the extension of the topological treatment to the timelike circular orbit. In this paper, we confirm that the angular momentum does not affect the asymptotic behavior of the constructed vector with its zero points denoting the equatorial timelike circular orbits. As a result, a well-behaved topology to characterize the equatorial timelike circular orbits can be constructed. Our study shows that the total topological number of the timelike circular orbits vanishes for a generic black hole, which is dependent of the energy of the particle. Significantly, it reveals that if there the timelike circular orbits exist, they always come in pairs for fixed angular momentum. Meanwhile, the stable and unstable timelike circular orbits have positive or negative winding number. Of particular interest is that the marginally stable circular orbit corresponds to the bifurcation point of the zero point of the constructed vector. Moreover, we also examine the case when the particle energy acts as the control parameter. It is shown that there will be topological phase transition when the value of the particle energy is one. Below this value, the timelike circular orbits always come in pairs for fixed energy. Otherwise, we will have one more unstable timelike circular orbit. Such a topological phase transition actually measures whether there are bounded orbits. We further apply the treatment to the Kerr black hole. All the results given in a generic black hole background are exactly reproduced. These strongly indicate that our topological approach can be generalized to the equatorial timelike circular orbits.
\end{abstract}



\maketitle

\section{Introduction}

Topology provides a powerful tool to understand the light rings (LRs) around compact objects and black holes. It was shown that LRs can be well-explored by calculating the topological charge while ignoring where the LRs are located \cite{Cunhaa,Cunhab}. Due to the fact that the LRs are closely related to the ringdown stage of a black hole merger and formation of black hole shadow, further study of the topology approach may peek into the modern observations of the LIGO/Virgo Collaboration \cite{Abbott} and EHT Collaboration \cite{Akiyama1,Akiyama2}.

In 2017, Cunha, Berti, and Herdeiro \cite{Cunhaa} constructed a special vector by using the null geodesics and found that the LRs are exactly at the zero points of the vector. Then in the vector space, the winding number of a zero point can provide us with some interesting topological results. Meanwhile, summing the winding numbers of all the zero points, one can obtain the topological charge for the LRs. Via such a topological approach, they originally proposed that under a stationary and axisymmetric equilibrium ultracompact object, the topological charge of the LRs always vanishes. This indicates that if the LRs exist, they must come in pairs for such ultracompact objects. Moreover, if one of the LRs is stable, the other one must be unstable \cite {Crispino,Hoda}.

Shortly afterwards, the study was generalized to the black hole backgrounds \cite{Cunhab}. The result suggests that for a stationary, axisymmetric, asymptotically flat black hole spacetime in four dimensions, it possesses at least one standard LR outside the horizon for each rotation sense. Other related issues including the stability of angular motion were discussed in Refs. \cite{Guo,Junior,Ghosh}. In particular, it was found that there exists a topological phase transition for the dilatonic Melvin spacetime and Schwarzschilld-dilatonic-Melvin black hole at a certain value of the dilatonic parameter \cite{JuniorYang}.

Combining this with Duan's topological current $\phi$-mapping theory \cite{Duan}, we also applied the study to the nonrotating black hole case \cite{Wei2019}. The inner structure of the topological charge, as well as the bifurcation phenomena was investigated. For the Dyonic black hole solution \cite{Lu}, the previous study showed that there exist several photon spheres (PSs), which are the counterpart of LRs in spherically symmetric spacetime. This provides us with a good opportunity to test the nontrivial topological argument. After calculating the topological charge, we found that its value is always -1 independent of other parameters. This implies that the standard PSs are always one more than the exotic ones. So the topological approach holds even when multi PSs are presented. The black hole spin will also not affect the value of the topological charge.

Inspiring by the study, we proposed a topological approach for the black hole thermodynamics \cite{Wei2022}. Starting with the first law of black hole thermodynamics, we constructed a vector. The critical point of the phase transition was found to exactly locate at the zero point of the vector. By making use of the Duan's topological current $\phi$-mapping theory, we endowed each critical point with a winding number. Summing the winding numbers for all the critical points, we shall obtain the topological charge, which remains unchanged for the system. Unexpectedly, we found that besides the conventional critical point of negative winding number, there is another novel critical point with positive winding number. Combining with the black hole thermodynamics, we pointed out that only the conventional critical point acts as an indicator of the first order black hole phase transition. Such a topological study was also extended to the Gauss-Bonnet black hole \cite{Yerra}, where three critical points appear. Nevertheless, the total topological charge remains unchanged with the black hole electric charge.

Considering that the black hole solutions with the critical points are very limited, we introduced another topological approach by the free energy of the black hole system \cite{Weito}. Each actual black hole solution corresponds to the zero point of a constructed vector. Thus, it will be endowed with a winding number in the vector space. Taking the inverse Hawking temperature as the control parameter, the total topological number is universal and independent of the black hole temperature and other parameters. By taking advantage of such topological properties, different black hole systems are divided into different topological classes. For example, the Schwarzschild, Reissner-Nordstr\"{o}m, and Reissner-Nordstr\"{o}m-AdS black holes, respectively, have topological number -1, 0, 1, which indicates they belong in different topological classes. All these results indicate that the topological approach is a promising way to study the LRs and thermodynamic properties of the black hole systems.

Comparing with the photons, the massive particles can also go around the black hole along the timelike circular orbits (TCOs). In certain astronomical processes, such orbits serve as the endpoints of dynamical evolution due to the dissipative effects \cite{Peters,Rodrigo}. In particular, there is the innermost stable circular orbit (ISCO), below which the massive particles shall quickly plunge into the black hole. Therefore, the stable accretion disk can only be formed above this bound, and provide a light source to illuminate the black hole, which may cast potential features of the black hole shadow in the observation of EHT \cite{Akiyama1,Akiyama2}. For a particle falls towards the black hole at a large radius, it will pass through a series of TCOs by losing angular momentum, and converting its energy into radiation, and finally reaches the ISCO before it plunges into the black hole. In the extremal Kerr black hole background, a particle could convert its 42\% energy, or achieve a higher efficiency in compact bosonic objects \cite{Delgado}. Via this process, which leads to the possible fact that black holes could be the powerful luminosity sources to trigger the energetic astronomical phenomena in strong gravity regime.

On the other hand, very recently, the general properties of the TCOs have been explored in Refs. \cite{Delgado,Cardoso}. It was found that the stable and unstable LRs delimit the region of stable and unstable TCOs. However, the topological approach for the TCOs remains to be established despite the great success gained for the LRs. In the backgrounds of compact objects or black holes, the locations of LRs are only dependent of the parameters of the compact objects or black holes, while independent of the energy and angular momentum of the photon. This states that the LRs are intrinsic properties of the spacetime. However, not all the photons can surround the black hole along these LRs. Considering that these photons come from far away from the black hole, some of them will fall into and some of them will fly past it. Only those with appropriate energy and angular momentum will surround the black hole along the LRs. Moreover, if the background is fixed, the relation of the energy and angular momentum of the photon is linear.

Now we wonder whether a massive test particle with certain energy and angular momentum can feel the TCOs. Similar to the photon, there must be some conditions. As is well-known, the radius of the TCO depends both on the black hole parameters and the massive particles themselves. For a given TCO, the relation of the energy and angular momentum of these particles orbiting along the black hole is not linear anymore. Due to these facts, it seems that we cannot construct a well-behaved topological approach for the TCOs as expected, or it is impossible to obtain the corresponding topological information of the TCOs.

However, it is worthwhile pointing out that the greatest advantage of the topological approach is that it ignores the specific details, such as the relation of the energy and angular momentum, as well as the locations of the TCOs. Keeping this in mind, in order to establish a well-behaved topological approach, we just require that, in certain parameter regions, the asymptotic behavior of the constructed vector does not change. Motivated by this idea, we manage to construct a topological approach for the TCOs for the first time. We shall show that when taking the angular momentum as the control parameter, the energy of the massive particle does not change the asymptotic behavior of the considered vector, indicating a well-behaved topology. In such approach, the marginally stable circular orbit (MSCO) acts as a bifurcation point of the zero points of the vector. More importantly, we find that the TCOs with the same angular momentum (or energy less than one) always come in pairs for the Kerr-like black holes.

The present paper is organized as follows. In Sec. \ref{Circular}, we start with the geodesics and construct a vector with its zero points corresponding to the TCOs. The asymptotic behavior of the vector is examined when we take the angular momentum as the control parameter, which means that the TCOs evolve with the angular momentum. In Sec. \ref{Topology}, we calculate the global and local topological charges for the TCOs. Meanwhile, the topological property of the MSCO is also discussed. We then apply this approach to the Kerr black hole in Sec. \ref{Kerr} as an example. Adopting energy as the control parameter, we investigate the corresponding topology in Sec. \ref{Control}. Finally, we summarize and discuss our results in Sec. \ref{Conclusion}.

\section{Circular geodesics and effective potential}
\label{Circular}

We consider a four-dimensional stationary, axisymmetric, asymptotically-flat black hole spacetime possessing two Killing vectors, $\xi^{\mu}=(\partial_{t})^{\mu}$ and $\psi^{\mu}=(\partial_{\varphi})^{\mu}$. Moreover, we assume the metric is circular, i.e., $\xi^{\mu}R_{\mu}^{[\nu}\xi^{\rho}\psi^{\sigma]}=\psi^{\mu}R_{\mu}^{[\nu}\xi^{\rho}\psi^{\sigma]}=0$, and has a north-south $\mathcal{Z}_2$ symmetry. As a result, the circular orbit motion can only be on the equatorial plane. Then under these conditions, the most generic metric reads
\begin{eqnarray}
 ds^2=g_{tt}(r, \theta)dt^2+g_{rr}(r, \theta)dr^2+g_{\theta\theta}(r, \theta)d\theta^2+g_{\varphi\varphi}d\varphi^2+2g_{t\varphi}dtd\varphi,\label{metrica}
\end{eqnarray}
assuming a signature (-, +, +, +). In addition to the Kerr black hole, most known black holes can also be described by this metric. For a black hole, we suppose that its horizon is located. at a constant (positive) radial coordinate $r=r_{\text{h}}$. So in the exterior region of the black hole, one should have $r_{\text{h}}<r<\infty$. We also have $g_{rr}>0$, $g_{\theta\theta}>0$, and $g_{\varphi\varphi}>0$ outside the horizon. The angular coordinates $\theta\in[0, \pi]$ and $\varphi\in[0, 2\pi]$. According to the $\mathcal{Z}_2$ symmetry, the equatorial plane is at $\theta=\pi/2$. Although we only concern the equatorial TCOs, here we still expect to leave $\theta$ as a free variable. Considering the condition det$(-g)>0$, one easily obtains
\begin{eqnarray}
 B(r, \theta)\equiv g_{t\varphi}^2-g_{tt}g_{\varphi\varphi}>0.\label{BBB}
\end{eqnarray}
At the horizon, we have $B(r_{\text{h}}, \theta)=0$. Under the metric (\ref{metrica}), the motion of a test particle can be described by the Lagrangian
\begin{eqnarray}
 \mathcal{L}=\frac{1}{2}g_{\mu\nu}\dot{x}^{\mu}\dot{x}^{\nu}=-\frac{1}{2}\mu^2,
\end{eqnarray}
where the dots denote the derivative with respect to an affine parameter, and $\mu^2=$1, 0, -1 are for the timelike, null, and spacelike geodesics, respectively. The conjugate momenta can be calculated as
\begin{eqnarray}
 \pi_{\mu}=\frac{\partial \mathcal{L}}{\partial \dot{x}^{\mu}}=g_{\mu\nu}\dot{x}^{\nu}.
\end{eqnarray}
Therefore, the Hamiltonian for the test particle reads
\begin{eqnarray}
 \mathcal{H}&=&\pi_{\mu}\dot{x}^{\mu}-\mathcal{L}\nonumber\\
 &=&\frac{1}{2}\left(g_{tt}\dot{t}^{2}+g_{rr}\dot{r}^{2}+g_{\theta\theta}\dot{\theta}^2
 +g_{\varphi\varphi}\dot{\varphi}^2+2g_{t\varphi}\dot{t}\dot{\varphi}\right)=-\frac{1}{2}\mu^2.
\end{eqnarray}
Rearranging it, we get
\begin{eqnarray}
 g_{rr}\dot{r}^{2}+g_{\theta\theta}\dot{\theta}^2+g_{tt}\dot{t}^{2}+2g_{t\varphi}\dot{t}\dot{\varphi}
 +g_{\varphi\varphi}\dot{\varphi}^2+\mu^2=0.\label{moot}
\end{eqnarray}
Following the treatment for the null geodesics \cite{Grover,Herdeiro}, we denote the kinetic term $\mathcal{K}$ and potential term $\mathcal{V}$ as
\begin{eqnarray}
 \mathcal{K}&=&g_{rr}\dot{r}^{2}+g_{\theta\theta}\dot{\theta}^2,\\
 \mathcal{V}&=&g_{tt}\dot{t}^{2}+2g_{t\varphi}\dot{t}\dot{\varphi}
 +g_{\varphi\varphi}\dot{\varphi}^2+\mu^2.\label{potent}
\end{eqnarray}
Therefore, the motion of the test particle (\ref{moot}) turns to
\begin{eqnarray}
 \mathcal{K}+ \mathcal{V}=0.
\end{eqnarray}
Note that the kinetic term $\mathcal{K}\geq0$, and the inequality is only saturated at $\dot{r}=\dot{\theta}=0$. Then the motion of the particle is completely governed by the effective potential $\mathcal{V}$. On the other hand, these two Killing vectors, $\xi^{\mu}$ and $\psi^{\mu}$, are related to two constants, the energy and orbital angular momentum of the test particle, along each geodesic
\begin{eqnarray}
 -E&=&g_{\mu\nu}u^{\mu}\xi^{\nu}=g_{tt}\dot{t}+g_{t\varphi}\dot{\varphi},\\
 l&=&g_{\mu\nu}u^{\mu}\psi^{\nu}=g_{t\varphi}\dot{t}+g_{\varphi\varphi}\dot{\varphi},
\end{eqnarray}
where $u^{\mu}$ denotes the tangent vector of the geodesics. Solving the above two equations, one can obtain
\begin{eqnarray}
 \dot{t}&=&\frac{1}{B}\left(Eg_{\varphi\varphi}+lg_{t\varphi}\right),\\
 \dot{\varphi}&=&-\frac{1}{B}\left(Eg_{t\varphi}+lg_{tt}\right).
\end{eqnarray}
Plugging them into the potential (\ref{potent}), one has
\begin{eqnarray}
 \mathcal{V}=-\frac{1}{B}\left(E^2g_{\varphi\varphi}+2Elg_{t\varphi}+l^2g_{tt}\right)+\mu^2.
 \label{poa}
\end{eqnarray}
When $\mu^2=0$, this exactly reduces to the potential of a photon \cite{Grover,Herdeiro}. Similarly, we can reexpress this potential as
\begin{eqnarray}
 \mathcal{V}=-\frac{l^2g_{\varphi\varphi}}{B}\left(\frac{E}{l}-H_{+}\right)\left(\frac{E}{l}-H_{-}\right)+\mu^2,
 \label{pot}
\end{eqnarray}
where $H_{\pm}$ are only dependent of the black hole metric and are given by
\begin{eqnarray}
 H_{\pm}=\frac{-g_{t\varphi}\pm\sqrt{B}}{g_{\varphi\varphi}}.
\end{eqnarray}
For a photon with $\mu^2=0$, it is quite clear that the LR related with $\mathcal{V}=\partial_{r}\mathcal{V}=0$ satisfies \cite{Grover,Herdeiro}
\begin{eqnarray}
 \frac{E}{l}&=&H_{\pm},\label{hha}\\
 \frac{\partial H_{\pm}}{\partial r}&=&0.\label{hhh}
\end{eqnarray}
This is an amazing result that the radius $r_{LR}$ of the LR (or the photon sphere for the static spherically black hole) of the black hole can be solved by Eq. (\ref{hhh}) and is independent the properties of the photon, indicating that it is an intrinsic structure of the black hole just like the horizon. Then plugging it into (\ref{hha}), one shall get $E/l=H_{\pm}(r_{LR})$ for these photons orbiting along the LR. In Ref. \cite{Cunhab}, by defining a new vector $v=((\partial_r H_{\pm})/\sqrt{g_{rr}}, (\partial_\theta H_{\pm})/\sqrt{g_{\theta\theta}})$, one finds the location of the LR is exactly at the zero point of $v$ from (\ref{hhh}). As a result, by calculating the winding number in the vector space, we can obtain some properties of the LR. In particular, considering all the regions of ($r$, $\theta$), the global property of the LR in a black hole background will be obtained by examining the asymptotic behaviors of $v$ near the boundary. This treatment mainly takes advantage of the result that $H_{\pm}$ are independent of the energy and angular momentum of the photon. However, due to the presence of the extra term $\mu^2$ in (\ref{pot}), the case becomes unclear.

In order to establish the topology for the TCO, we reexpress the effective potential (\ref{poa}) in the following form
\begin{eqnarray}
 \mathcal{V}=-\frac{g_{\varphi\varphi}}{B}(E-e_{1})(E-e_{2}),\label{ppotent}
\end{eqnarray}
where
\begin{eqnarray}
 e_{1}=\frac{-lg_{t\varphi}+\sqrt{B}\sqrt{l^2+\mu^2g_{\varphi\varphi}}}
    {g\varphi\varphi},\label{ee1}\\
 e_{2}=\frac{-lg_{t\varphi}-\sqrt{B}\sqrt{l^2+\mu^2g_{\varphi\varphi}}}
    {g\varphi\varphi}.
\end{eqnarray}
Obviously, $\mathcal{V}=0$ leads to the energy of the test particle $E=e_{1,2}$. Considering the condition (\ref{BBB}) and $g_{\varphi\varphi}>0$, it is easy to obtain $e_2<g_{t\varphi}(|l|-l)/g_{\varphi\varphi}$. For $l\geq0$, it is easy to obtain $e_2<0$, and for $l<0$, $e_2<-2lg_{t\varphi}/g_{\varphi\varphi}<0$ due to negative $g_{t\varphi}$. Therefore, it is notable that for any value of $l$, $e_{2}$ takes a negative value. As a result, we abandon it. Then, the conditions $\mathcal{V}=\partial_{r}\mathcal{V}=0$ for determining the TCO become
\begin{eqnarray}
 E&=&e_{1},\label{pace} \\
 \partial_{r}e_{1}&=&0.\label{pacl}
\end{eqnarray}
Quite different from the LR, $\partial_{r}e_{1}$ here not only depends on $r$, but also on the angular momentum $l$ of the test particle, so it seems that we cannot construct the topology for the TCO as expected.

However, we would like to propose a new viewpoint while is still similar to that for the photon. For each given angular momentum $l$, we can obtain the radius of the TCO by solving (\ref{pacl}). Whether the equation has solution or not, the angular momentum $l$ of the test particle does not affect our treatment. After obtaining the radius of the TCO, we can substitute it into (\ref{pace}), and then the energy of the test particle will be obtained.

In order to give a global topology, we only require that the values of the angular momentum do not change the asymptotic behavior of $\partial_{r}e_{1}$ at the boundary of the ($r$, $\theta$) plane. Aiming at it, we define a new vector $\phi$=($\phi^r$, $\phi^\theta$) by analog to Ref. \cite{Cunhab} for the photon,
\begin{eqnarray}
 \phi^r=\frac{\partial_{r} e_{1}}{\sqrt{g_{rr}}},\quad
 \phi^\theta=\frac{\partial_{\theta} e_{1}}{\sqrt{g_{\theta\theta}}},\label{vectc}
\end{eqnarray}
in a flat vector space. In order to be consistent with the notations of Ref. \cite{Duan}, we write these indices with up ones. Obviously, $\phi=0$ corresponds to the TCO and $\theta=\pi/2$ as expected. Next, we turn to the asymptotic behavior of $\phi$ at the boundary of the ($r$, $\theta$) plane; this shall further confirm that the construction of the topology for the TCO is meaningful.

For our case, we should examine four asymptotic behaviors of $\phi$ near the horizon $r\rightarrow r_{\text{h}}$, infinity $r\rightarrow\infty$, and $\theta\rightarrow 0$, $\pi$.

i) \emph{Axis limits}. Here we shall consider the axis limits of the vector $\phi$. In order to clearly show them, we here follow Ref. \cite{Cunhab} and give a detailed calculation. Defining a local coordinate $\rho=\sqrt{g_{\varphi\varphi}}$, one can find when $\theta\rightarrow0$ or $\pi$, $d\rho/d\theta$ is positive or negative. Since the metric functions are independent of the particles, one has the same result as that of Ref. \cite{Carot} by expanding the metric function at small $\rho$
\begin{eqnarray}
 &&g_{tt}\sim g_{tt}^{0}+\mathcal{O}(\rho),\quad
 g_{\rho\rho}\sim g_{\rho\rho}^{0}+\mathcal{O}(\rho),\\
 &&g_{\varphi\varphi}=\rho^2,\quad
 g_{t\varphi}\sim b_{0}\rho^{n}+\mathcal{O}(\rho^{n+1}),
\end{eqnarray}
where $ g_{tt}^{0}$, $g_{\rho\rho}^{0}$, $b_{0}$ are constants, and $n\in N$. Assuming $C^2$ smoothness and regularity close to the axis, one shall have $n\geq2$ \cite{Carot}. Plugging them into Eq. (\ref{ee1}), we obtain
\begin{eqnarray}
 e_{1}\sim\frac{\sqrt{-g_{tt}^{0}l^2\rho^2+\sqrt{-g_{tt}^{0}}\mu^2\rho^3}}{\rho^2}.
\end{eqnarray}
At zero order in $\rho$, one has $g_{\rho\rho}d\rho^2\sim g_{\theta\theta}d\theta^2$, and thus
\begin{eqnarray}
 \phi^{\theta}\sim \text{sign}\left(\frac{d\rho}{d\theta}\right)\partial_{\rho}e_{1}
  \sim-\text{sign}\left(\frac{d\rho}{d\theta}\right)\frac{|l|\sqrt{-g_{tt}^{0}}}{\rho^2},\nonumber\\
 \phi^{\theta}\sim \text{sign}\left(\frac{d\rho}{d\theta}\right)\partial_{\rho}e_{1}
  \sim-\text{sign}\left(\frac{d\rho}{d\theta}\right)\frac{(-g_{tt}^{0})^{\frac{1}{4}}\mu^2}{2\rho^{\frac{3}{2}}},
\end{eqnarray}
for nonvanishing and vanishing $l$, respectively. Therefore, it is clear that $\phi^{\theta}$ diverges at $\rho\to 0$ and the angular momentum do not affect the behavior of $\phi^{\theta}$ at small $\rho$. Further combining with $\phi^{r}\sim \rho^{-1}$, we have $\phi^{\theta}\gg\phi^{r}$. This indicates that the direction of the vector is vertical near $\theta$=0 and $\pi$. Moreover, considering the fact that $d\rho/d\theta$ is positive and negative as $\theta\rightarrow$0 and $\pi$, we find the vector is outward at $\theta$=0 and $\pi$.

ii) \emph{Horizon limit}. Here we would like to consider the behavior of the vector $\phi$ when $r\to r_{h}$. As discussed in Ref. \cite{Medved}, we can set a local radial coordinate $x$ such that $g_{xx}=1$, and $x=0$ at the black hole horizon. Near the horizon, one can obtain these metric functions \cite{Medved}
\begin{eqnarray}
 \omega\simeq\omega_{H}+\mathcal{O}(x^2),\quad
 g_{\varphi\varphi}\simeq g_{\varphi\varphi}^{H}+\mathcal{O}(x^2),\\
 g_{t\varphi}\simeq-\omega_{H} g_{\varphi\varphi}^{H},\quad
 B\simeq g_{\varphi\varphi}^{H}\kappa^{2}x^{2},\quad
 g_{tt}\simeq \omega_{H}^2g_{\varphi\varphi}^{H},
\end{eqnarray}
where $\omega=-g_{t\varphi}/g_{\varphi\varphi}$ and $\kappa$ denotes the surface gravity of the black hole. $g_{\varphi\varphi}^{H}$ and $\omega_{H}$ are zero-order terms of the expansions in $x$. Then
\begin{eqnarray}
 \partial_{x}e_{1}\simeq \kappa\sqrt{\frac{l^2}{g_{\varphi\varphi}^{H}}+\mu^2}+\mathcal{O}(x).
\end{eqnarray}
By using $(1/\sqrt{g_{xx}})(\partial/\partial x)=(1/\sqrt{g_{rr}})(\partial/\partial r)$, one shall obtain
\begin{eqnarray}
 \phi^{r}=\frac{\partial_{r}e_{1}}{\sqrt{g_{rr}}}\simeq \kappa\sqrt{\frac{l^2}{g_{\varphi\varphi}^{H}}+\mu^2}.\label{kaidao}
\end{eqnarray}
For a nonextremal black hole with $\kappa>0$, we see that $\phi^{r}$ takes a positive value. In order to determine the direction of the vector $\phi$ at the horizon, one needs to calculate $\phi^{\theta}$. However, no matter whether $\phi^{\theta}$ is positive or negative, we can claim that the direction of the vector $\phi$ must be to the right, with at most an inclination.

iii) \emph{Asymptotic limit}. Here let us consider the asymptotic behavior of the vector at infinity. Here we require the spacetime tends to a flat one in
standard spherical coordinates, namely
\begin{eqnarray}
 g_{t\varphi}&\simeq&-\mathcal{O}\left(\frac{1}{r}\right),\quad
 g_{\varphi\varphi}\simeq r^{2}\sin^{2}\theta+\mathcal{O}\left(1\right),\nonumber\\
 g_{tt}&\simeq& -1+\frac{2M}{r}+\mathcal{O}\left(\frac{1}{r^2}\right),\quad
 g_{rr}\simeq 1+\frac{2M}{r}++\mathcal{O}\left(\frac{1}{r^2}\right).
\end{eqnarray}
Parameter $M$ is a positive constant and can be treated as the black hole mass in the asymptotically flat spacetime. Therefore, we have
\begin{eqnarray}
 \phi^{r}\simeq\frac{M\sqrt{\mu^2}}{r^{2}}+\mathcal{O}\left(\frac{1}{r^{3}}\right),
\end{eqnarray}
on the equatorial plane by taking $\theta=\pi/2$, which is small but a positive value. As a result, by ignoring the specific value of $\phi^{\theta}$, the direction of the vector $\phi$ is to the right similar to that in the horizon limit.

In summary, we see that the angular momentum $l$ of the test massive particle do not affect the behavior of the vector $\phi$ at the boundary of the ($r$, $\theta$) plane in our above approach. This result strongly implies that we can construct the topology for the TCOs of the massive particle in a black hole background.

\section{Global and local topologies}
\label{Topology}

Based on the asymptotic behavior of $\phi$ at the boundary of the ($r$, $\theta$) plane, we here investigate the global and local topological properties for the TCOs. Moreover, we also examine the topology for the MSCO and ISCO.

\subsection{Global property}

As we have shown that the TCOs locate exactly at the zero points of the vector $\phi$. Hence, we can calculate the topological number $W$ defined as \cite{Duan}
\begin{eqnarray}
 W=\int_{\Sigma}j^{0}d^{2}x
\end{eqnarray}
for a giving region $\Sigma$ in the ($r$, $\theta$) plane for the TCOs. Here $j^{0}$ is the zeroth component of the topological current
\begin{eqnarray}
 j^{\mu}=\frac{1}{2\pi}\epsilon^{\mu\nu\rho}\epsilon_{ab}
 \frac{\partial n^{a}}{\partial x^{\nu}}\frac{\partial n^{b}}{x^{\rho}},
\end{eqnarray}
where $x^{\mu}=(t, r, \theta)$, and the unit vector $n^{a}$=($n^{r}$, $n^{\theta}$)=($\phi^{r}/|\phi|$, $\phi^{\theta}/|\phi|$). Here, $x^0=t$ denotes the time control parameter. One can study the evolution of the zero points with that parameter as expected. On the other hand, other parameters of the black hole or particle can also be chosen as the control parameter. For a good definition, we require that the zero points cannot reach the boundary of the ($r$, $\theta$) plane at finite value of the control parameter. In the following study, we mainly treat the angular momentum $l$ of the particle as a control parameter. The brief discussion on choosing energy will also be given.

After some calculations, one arrives at \cite{Duan}
\begin{eqnarray}
 W=\sum_{n=1}^{N}w_{n},
\end{eqnarray}
which is the sum of the winding number $w$ for all the zero points on the region $\Sigma$. On the other hand, we can also obtain the topological charge for a region $\Sigma$ by counting the number of the loops that the vector makes in the vector $\phi$ space when $x^{\mu}$ moves along the closed curve $\partial\Sigma$ in a counterclockwise direction,
\begin{eqnarray}
 W=\frac{1}{2\pi}\oint_{C}d\Omega=\frac{1}{2\pi}\oint_{C}\epsilon_{ab}n^{a}dn^{b},
\end{eqnarray}
where $\Omega$ denotes the change of the direction of the vector.

\begin{figure}
\includegraphics[width=7cm]{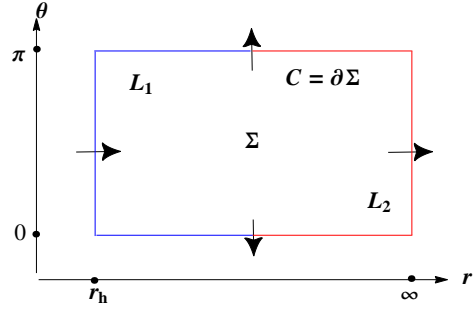}
\caption{Representation of the contour $C=\partial\Sigma=L_1\cup L_2$ (which encloses all the parameter region $\Sigma$) on the
($r$, $\theta$) plane. $L_1$ and $L_2$ are the left and right parts of the contour marked in blue and red colors, respectively. Black arrows indicate the approximate directions of the vector $\phi$ in the boundaries.}\label{rtheta}
\end{figure}
%%%%%%%%%%%%

Considering that the curve $C=\partial\Sigma$ encloses the entire parameter region, one will obtain the global property of the topology. Here we can divide the closed curve into two parts $L_{1}$ and $L_{2}$ such that $C=L_{1}\cup L_{2}$. As shown in Fig. \ref{rtheta}, we can see that the direction of the vector are outward at $\theta=0$ and $\pi$, and is to the right at $r=r_{\text{h}}$ and $\infty$ by ignoring the possible inclination. According to the asymptotic behaviors of $\phi$, we easily get
\begin{eqnarray}
 W&=&\frac{1}{2\pi}\int_{L_1}d\Omega+\frac{1}{2\pi}\int_{L_2}d\Omega\nonumber\\
 &=&\frac{1}{2\pi}\times(-\pi)+\frac{1}{2\pi}\times(\pi)\nonumber\\
 &=&0.
\end{eqnarray}
This states that the total topological number vanishes for the TCOs. Significantly, it is universal result and is independent of the angular momentum of the massive test particle, as well as the black hole spin. This is an amazing result, and based on which, we can conclude that the TCOs with the fixed angular momentum always come in pairs for the four-dimensional stationary, axisymmetric, asymptotically flat black hole. This property is quite different from that of the LR, whose topological number is -1 indicating that at least one one LR exists.

\begin{figure}
\center{\subfigure[]{\label{Inwards}
\includegraphics[width=6cm]{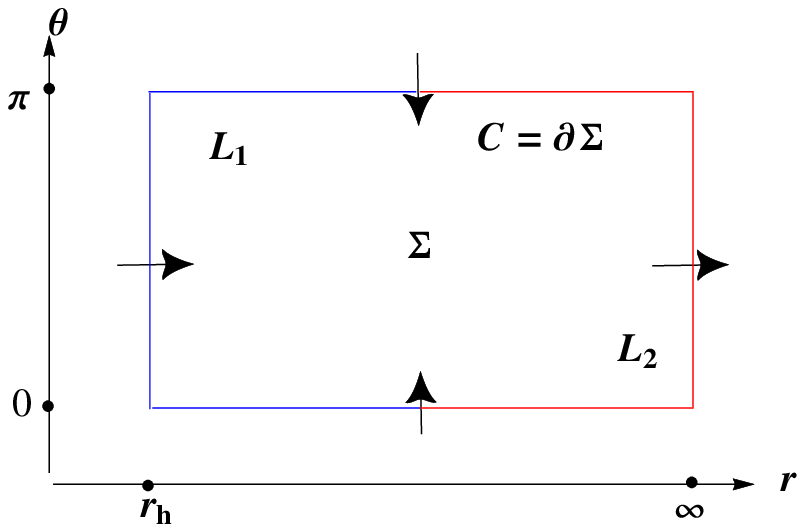}}
\subfigure[]{\label{Upup}
\includegraphics[width=6cm]{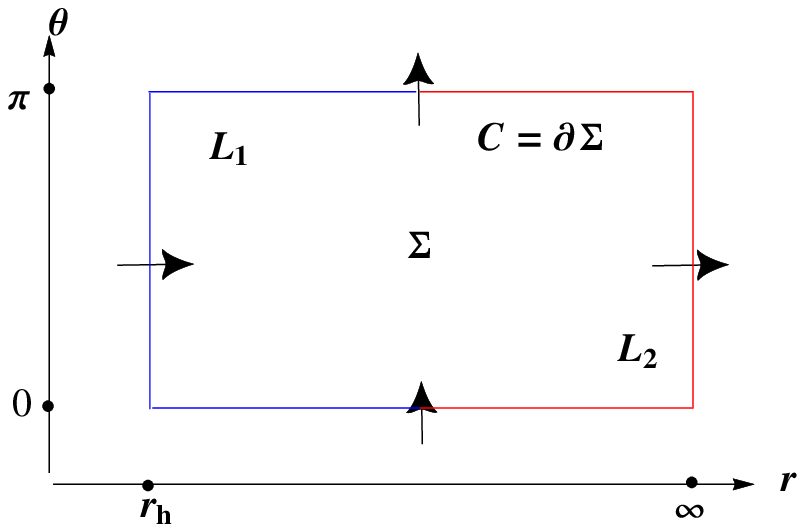}}\\
\subfigure[]{\label{DownDown}
\includegraphics[width=6cm]{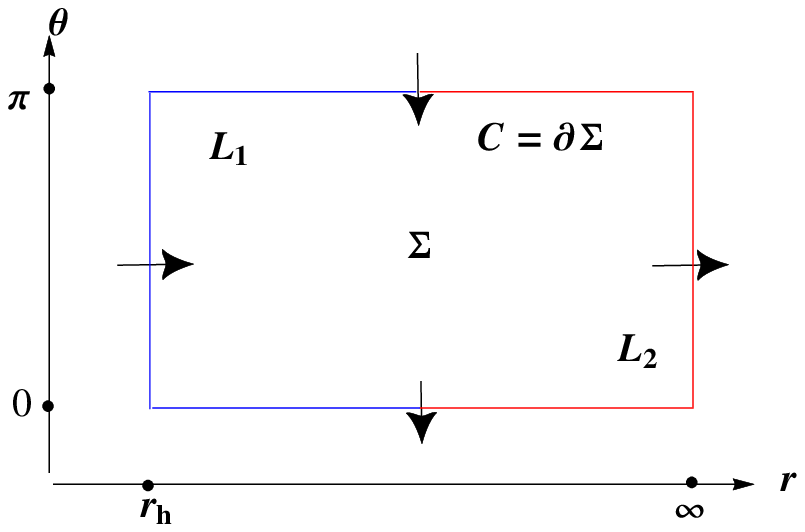}}}
\caption{Different patterns for the direction of the vector $\phi$ at $\theta$=0 and $\pi$.}\label{ppDownDown}
\end{figure}

As shown above, the total topological number for the TCOs vanishes where the direction of the vector $\phi$ is outwards at $\theta$=0 and $\pi$. However, even under the $\mathcal{Z}_2$ symmetry, one may wonder whether the topological number still vanishes if the direction of the vector $\phi$ is inwards at $\theta$=0 and $\pi$, see Fig. \ref{Inwards}. After considering the change of the direction of the vector $\phi$, we still obtain $W=0$. On the other hand, it is valuable to examine the total topological number if we abandon the $\mathcal{Z}_2$ symmetry. We further exhibit these two possible cases in Figs. \ref{Upup} and \ref{DownDown}, where the $\mathcal{Z}_2$ symmetry is broken as expected. Surprisingly, we still have
\begin{eqnarray}
 W=0,
\end{eqnarray}
for both the cases without $\mathcal{Z}_2$ symmetry.

In short, we find that the result $W=0$ always holds for the case with or without $\mathcal{Z}_2$ symmetry.

\subsection{Local property and stability}

As shown above, we confirm that the global topological number of the TCOs is zero. So they either disappear or appear in pairs. However, quite different from the LR which is the intrinsic structure of the spacetime, TCOs are closely dependent of the test particle. For some certain values of the energy and angular momentum, the TCOs may be present, while for other values, they may be absent. In this subsection, we focus on the local topological charge and the stability for the TCOs.

\begin{figure}
\includegraphics[width=7cm]{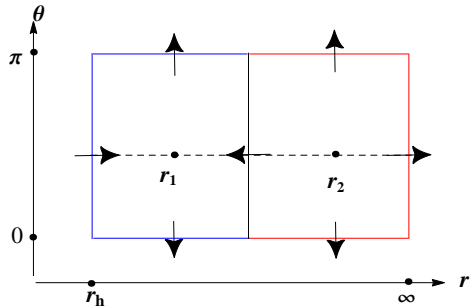}
\caption{Representation of the direction of the vector $\phi$ along the contours enclosing two zero points at $r_{1}$ and $r_{2}$.}\label{tbl}
\end{figure}

For each TCO at a given radius $r_{\text t}$, the energy and angular momentum are not independent of each other, and they must satisfy the conditions (\ref{pace}) and (\ref{pacl}). From (\ref{pacl}), one can solve the angular momentum as \cite{Delgado}
\begin{eqnarray}
 l_{\pm}=\frac{g_{t\varphi}+g_{\varphi\varphi}\Omega_{\pm}}{\sqrt{\beta_{\pm}}}\bigg|_{r_{\text t}},
 \label{lll}
\end{eqnarray}
where
\begin{eqnarray}
 \Omega_{\pm}&=&\frac{-g_{t\varphi}'\pm\sqrt{C}}{g_{\varphi\varphi}'},\\
 \beta_{\pm}&=&-g_{tt}-2g_{t\varphi}\Omega_{\pm}-g_{\varphi\varphi}\Omega_{\pm}^{2},\\
 C&=&(g_{t\varphi}')^2-g_{tt}'g_{\varphi\varphi}',\label{ccg}
\end{eqnarray}
with the prime denoting the derivative with respect to the radial coordinate.
Plugging the angular momentum $l$ into (\ref{pace}), one obtains the energy
\begin{eqnarray}
 E_{\pm}=-\frac{g_{tt}+g_{t\varphi}\Omega_{\pm}}{\sqrt{\beta_{\pm}}}\bigg|_{r_{\text t}}.
 \label{eee}
\end{eqnarray}
By making use of (\ref{ppotent}), we have
\begin{eqnarray}
 \frac{\partial^2 \mathcal{V}}{\partial r^2}=\frac{(e_{1}-e_{2})g_{\varphi\varphi}\sqrt{g_{rr}}}{B}\frac{\partial \phi^{r}}{\partial r},
\end{eqnarray}
for the TCOs. Since $e_2<0$, the coefficient is positive. Therefore, it is easy to know
\begin{eqnarray}
 \frac{\partial^2 \mathcal{V}}{\partial r^2}\bigg|_{r_{\text t}}>0(<0)\Rightarrow
 \frac{\partial \phi^{r}}{\partial r}\bigg|_{r_{\text t}}>0(<0).\label{vrr0}
\end{eqnarray}
Note that the stable and unstable TCOs correspond to positive and negative $\partial^2 \mathcal{V}/\partial r^2$, respectively. Via (\ref{vrr0}), we can also examine the radial stability of the TCOs by using $\phi^{r}$. In Fig. \ref{tbl} we show an example that there are two TCOs located at $r_1$ and $r_2$ (suppose $r_{1}<r_{2}$), respectively. Via the asymptotic behaviors of $\phi^{r}$ at horizon and infinity, we know $\phi^{r}$ is positive in the regions $(r_{\text{h}}, r_{1})\cup(r_{2}, +\infty)$, and negative in $(r_{1}, r_{2})$. So near $r=r_{1}$, $\partial \phi^{r}/\partial r$ is negative, while near $r=r_{2}$, $\partial \phi^{r}/\partial r$ is positive. This indicates that the TCO is unstable at $r_{1}$, while stable at $r_{2}$.

Here, we would like to briefly discuss the local stability in the angular direction. Here we only consider the equatorial TCOs with $\theta=\pi/2$. As we have shown in Sec. \ref{Circular}, the presence of $l$ does not change the axis limit of the vector at $\theta$=0 and $\pi$. Moreover, we can obtain the result that the local stability in the vertical direction of the TCOs is the same as that of the LRs. Then according to Refs. \cite{Cunhab,Guo}, it is easy to find that the TCOs are stable in the vertical direction.

%%%%%%%%%%%%

After addressing the stability of the TCOs, we expect to examine the value of the winding number for these two TCOs. Considering that the component $\phi^{r}$ of vector $\phi$ or $n^{r}$ of the unit vector $n$ changes its sign at $r_{1}$ and $r_{2}$, we also mark the direction of the vector in Fig. \ref{tbl}. Counting the change of the direction of the vector, one can easily get the winding number $w$=-1 and 1 for the zero points located at $r_{1}$ and $r_{2}$, respectively. So we can conclude that the unstable and stable TCOs correspond to negative and positive winding numbers. Nevertheless, the sum of the winding numbers vanishes.

\subsection{MSCO and bifurcation point}

We have shown above that positive or negative winding numbers correspond to stable or unstable TCO with $\partial^2 \mathcal{V}/\partial r^2>0$ or $<0$. Between these two cases, there is a critical case with $\partial^2 \mathcal{V}/\partial r^2=0$, which exactly corresponds to the MSCO or ISCO for the Kerr-like black holes.

As noted in Refs. \cite{Delgado,Raduba,Raduad}, MSCO and ISCO are two different concepts for (hairy) black hole or generic ultracompact objects. The MSCO denotes the stable TCO with the smallest radius, which can be continuously connected to spatial infinity by a set of stable TCOs. While ISCO denotes the stable TCO with the smallest radius among all the TCOs, which can be determined by $C=0$ (\ref{ccg}). For the Schwarzschild and Kerr black holes, the MSCO meets the ISCO; however, for other rotating black holes, such as the boson stars, Proca stars, and the hairy black holes, they are different. In order to avoid confusing, we here adopt the treatment of Ref. \cite{Delgado}.

In general, the MSCO is determined by
\begin{eqnarray}
 \mathcal{V}=0,\quad
 \frac{\partial \mathcal{V}}{\partial r}=0,\quad
 \frac{\partial^2 \mathcal{V}}{\partial r^2}=0.
\end{eqnarray}
The first condition is naturally satisfied with $E=e_{1}$, which is dependent of the angular momentum $l$ and radius $r_{\text t}$ of the TCOs. Meanwhile the radius $r_{\text t}$ can be obtained by solving the second equation. As expected, we denote its solution as $r=r_{\text t}(l)$. For different values of $l$, the TCOs have different orbit radii, so we can treat the angular momentum as a control parameter. After a simple calculation, the third condition $\partial^2\mathcal{V}/\partial r^2=0$ or $\partial\phi^{r}/\partial r=0$ turns to
\begin{eqnarray}
 \frac{dl}{dr_{\text t}}\bigg|_{(l^*, r_{\text t}^{*})}=0.
\end{eqnarray}
According to Duan's topological current theory, the point $(l^*, r_{\text t}^{*})$ related by the MSCO is a bifurcation point. Near the point, we can expand the angular momentum as
\begin{eqnarray}
 l-l^{*}=\frac{1}{2}\frac{d^2l}{dr_{\text t}^{2}}\bigg|_{(l^*, r_{\text t}^{*})}(r_{\text t}-r_{\text t}^{*})^2+\mathcal{O}(r_{\text t}-r_{\text t}^{*})^3.
\end{eqnarray}
Via this equation, one can obtain two branch solutions in the $r_{\text t}-l$ plane. When $\frac{d^2l}{dr_{\text t}^{2}}\big|_{(l^*, r_{\text t}^{*})}>0$, these two branch solutions are present for $l>l^{*}$, and otherwise, they are present for $l<l^{*}$. According to this property, the bifurcation point can be, respectively, divided into the generated and annihilated points \cite{Fu}.

Here, let us focus on the topological number of the MSCO. To clearly show it, we take the BP$_2$ branch described by the blue curve in Fig. \ref{CPara} as an example. For the large angular momentum, there are two TCOs, while no TCO for small one. This implies that BP$_2$ is a generated point. As we have shown above, for the fixed angular momentum, the TCOs with large and small radius are stable and unstable, and which have positive winding number $w$=1 and -1, respectively. Therefore, the upper branch has positive winding number, while the lower branch has negative winding number. Obviously, we can see that the branches with positive slope and negative slope, or equivalently with $\frac{\partial^2 \mathcal{V}}{\partial r^2}>0$ and $<0$ have $w$=1 and -1, respectively. Starting with a large angular momentum, the upper and lower branches have $\frac{\partial^2 \mathcal{V}}{\partial r^2}>0$ and $<0$. This property holds with the decrease of the angular momentum. However, when the point BP$_2$ is reached, such pattern changes. These two branches merge and only one MSCO with $\frac{\partial^2 \mathcal{V}}{\partial r^2}=0$ leaves. So one can regard the MSCO as one degenerated TCO. For any angular momentum slightly larger than that of BP$_2$, the MSCO will be split into two TCOs, one of which have $w$=1, while other one has $w$=-1. Considering that $\frac{\partial^2 \mathcal{V}}{\partial r^2}>0$ and $<0$, respectively, correspond to $w$=+1 and -1. The MSCO with $\frac{\partial^2 \mathcal{V}}{\partial r^2}=0$ will not take $w$=+1 or -1. Therefore, The MSCO must have $w$=0. This result is exactly consistent with the result of Ref. \cite{Duan}. Thus, we can say that the topological charge of the MSCO vanishes. It can also be further confirmed in Fig. \ref{KerrISCO} and Fig. \ref{KerrUTC} by calculating the change of the vector direction along the constructed closed loop.

\section{Timelike circular orbits and topology in Kerr black hole}
\label{Kerr}

In the previous section, we constructed the topology for the TCOs in a generic black hole background. In this section, we manage to apply it to the Kerr black hole and to confirm whether the above study is applicable. This shall uncover the detailed properties for a specific example.

\subsection{Effective potential}

The metric of the Kerr black hole is given by
\begin{eqnarray}
 ds^{2}=-\frac{\Delta}{\rho^{2}}\bigg(dt-a\sin^{2}\theta d\varphi\bigg)^{2}
        +\frac{\rho^{2}}{\Delta}dr^{2}+\rho^{2}d\theta^{2}\nonumber\\
        +\frac{\sin^{2}\theta}{\rho^{2}}\bigg(adt-(r^{2}+a^{2}) d\varphi\bigg)^{2},
\end{eqnarray}
where the metric functions read
\begin{eqnarray}
 \rho^{2}&=&r^{2}+a^{2}\cos^{2}\theta,\\
 \Delta&=&r^{2}-2Mr+a^{2}.
\end{eqnarray}
Solving $\Delta=0$, one can obtain the radii of the black hole horizons
\begin{eqnarray}
 r_{\pm}=M\pm\sqrt{M^2-a^2}.
\end{eqnarray}
Obviously, for a black hole, its spin must be $a/M\in(0, 1)$. From (\ref{ppotent}), the effective potential $\mathcal{V}$ can be obtained, and $e_{1,2}$ are given by
\begin{eqnarray}
 e_{1,2}=\frac{2 a l M r\pm\rho  \csc\theta \sqrt{\Delta
   \left(\left(a^2+r^2\right)^2\sin^2\theta+l^2 \rho^2-a^2\Delta\sin^4\theta\right)}}{\left(a^2+r^2\right)^2-a^2 \Delta  \sin
   ^2\theta},
\end{eqnarray}
where we have taken $\mu^2=1$. In the equatorial plane, they reduce to
\begin{eqnarray}
 e_{1,2}=\frac{2
   a l M r\pm r \sqrt{\Delta  \left(\left(a^2+r^2\right)^2+l^2 r^2-a^2 \Delta\right)}}{\left(a^2+r^2\right)^2-a^2 \Delta}.
\end{eqnarray}

\begin{figure}
\center{\subfigure[]{\label{Veffa}
\includegraphics[width=6cm]{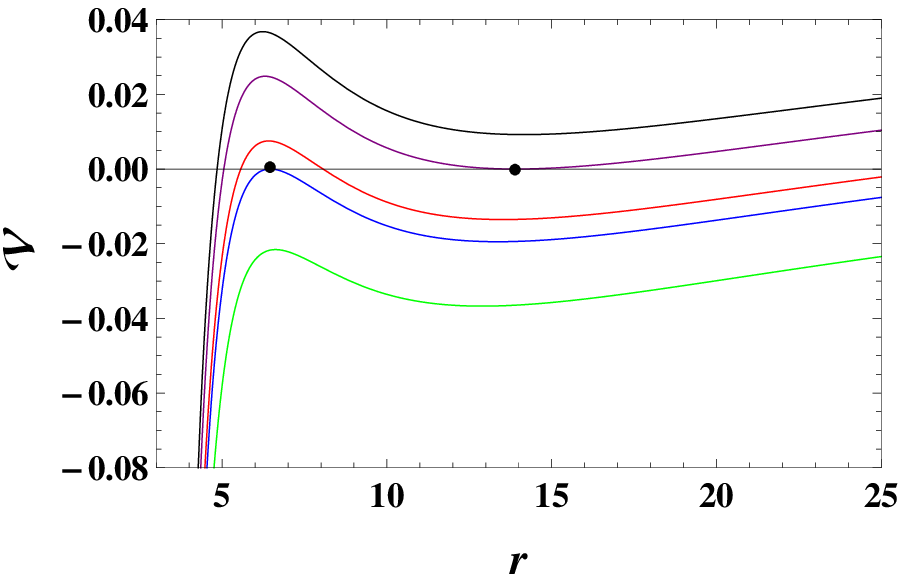}}
\subfigure[]{\label{Veffb}
\includegraphics[width=6cm]{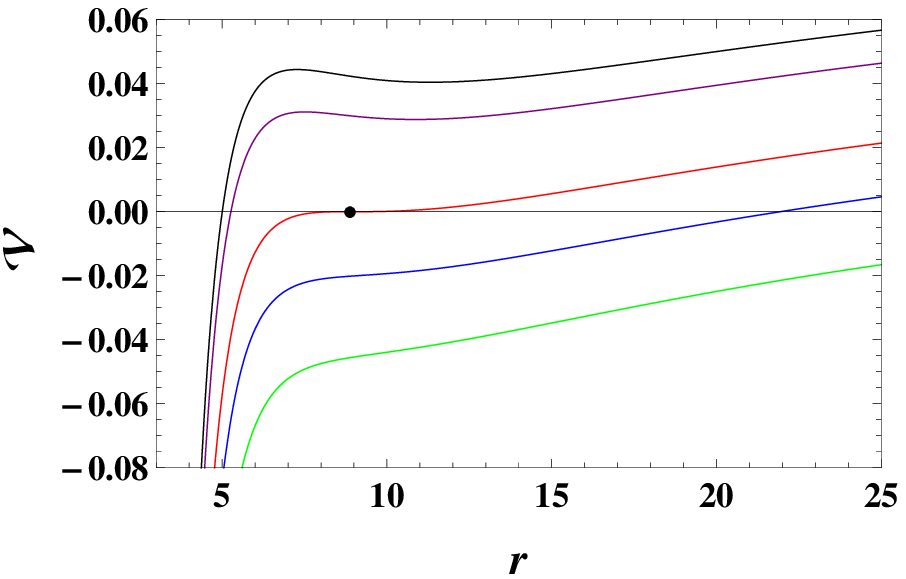}}\\
\subfigure[]{\label{Veffc}
\includegraphics[width=6cm]{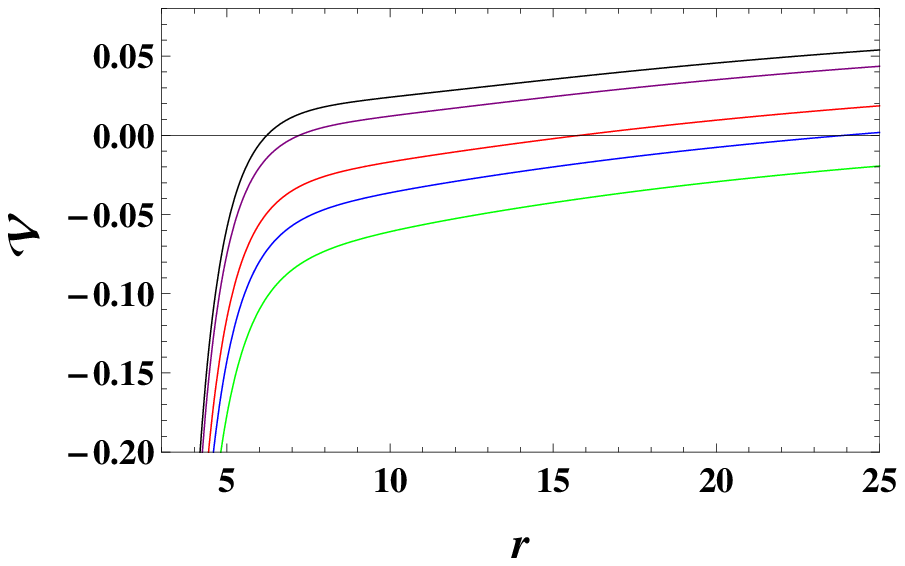}}
\subfigure[]{\label{Veffd}
\includegraphics[width=6cm]{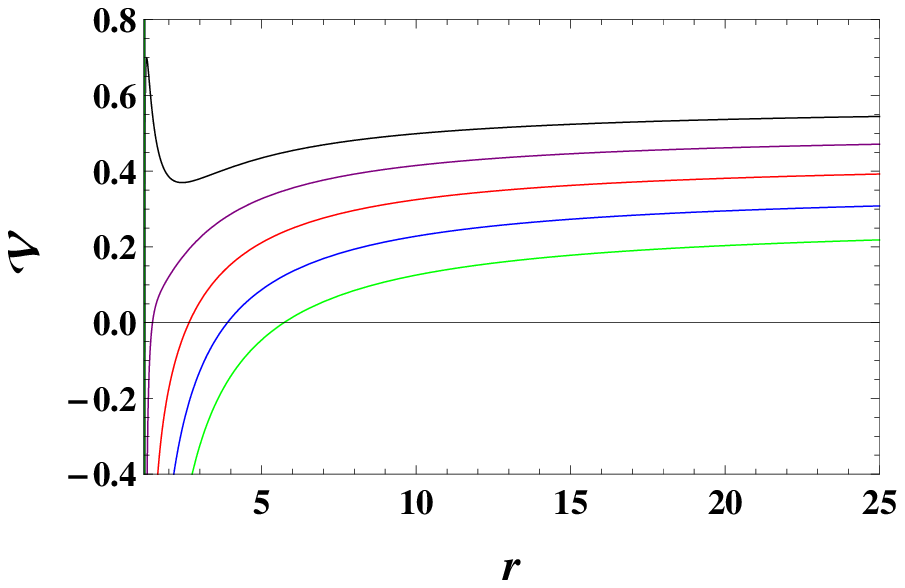}}\\
\subfigure[]{\label{Veffe}
\includegraphics[width=6cm]{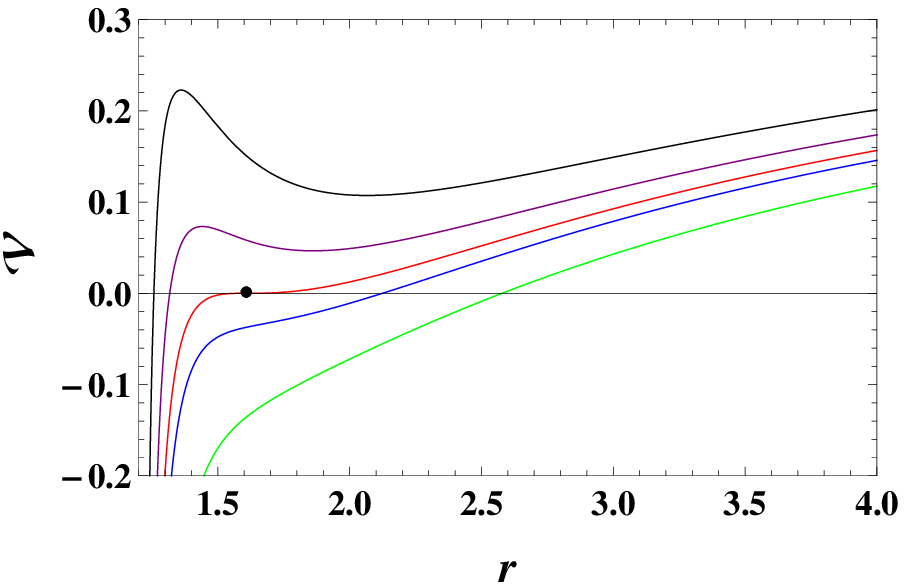}}
\subfigure[]{\label{Vefff}
\includegraphics[width=6cm]{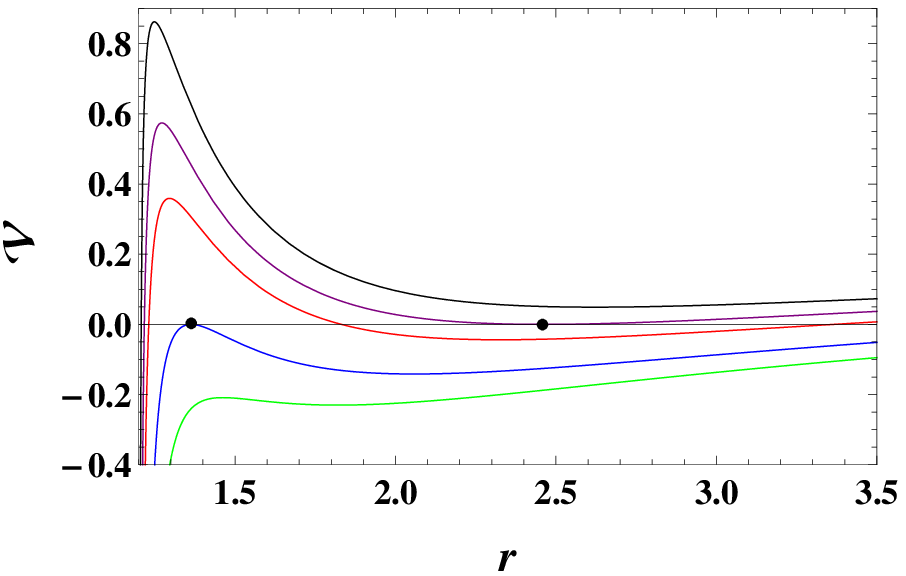}}}
\caption{Behavior of the effective potential $\mathcal{V}$ with $a$=0.98 and $M=1$. (a) $l$=-4.5. $E$=0.9650, 0.9691, 0.9750, 0.9776, and 0.9850 from top to bottom. (b) $l$=-4.2209. $E$=0.9450, 0.9500, 0.9620, 0.9700, and 0.9800 from top to bottom. (c) $l$=-4. $E$=0.9450, 0.9500, 0.9620, 0.9700, and 0.9800 from top to bottom. (d) $l$=1.5. $E$=0.65, 0.70, 0.75, 0.80, and 0.85 from top to bottom. (e) $l$=1.6827. $E$=0.75, 0.76, 0.7661, 0.77, and 0.78 from top to bottom. (f) $l$=1.9. $E$=0.81, 0.8210, 0.83, 0.8474, and 0.86 from top to bottom. Black dots denote these points satisfying $\mathcal{V}=\partial_{r}\mathcal{V}$=0. }\label{ppVefff}
\end{figure}

In order to show the characteristic behavior of the potential, we plot it in Fig. \ref{ppVefff} for the black hole with $a$=0.98 and $M=1$. When $l$=-4.5, we observe from Fig. \ref{Veffa} that there are two extremal points for fixed energy $E$. When $E$=0.9691 and 0.9776, there exists one point with $\mathcal{V}=\partial_{r}\mathcal{V}$=0 for each curve. It is obvious that these points are just the locations of TCOs, where the surrounding particles have zero radial velocity. So there are two TCOs for $l$=-4.5. Slightly increasing the angular momentum such that $l$=-4.2209, we can find from Fig. \ref{Veffb} that there are two extremal points for the small value of the energy. With the increase of the energy, these two extremal points coincide at the black dot with energy $E$=0.9620, and an extra condition $\partial_{r,r}\mathcal{V}=0$ is satisfied. Actually, the black dot denotes the MSCO as expected. In particular, it is worthwhile pointing out that such MSCO exactly meets the ISCO for the Kerr black hole. When the energy is beyond 0.9620, no extremal point will be present. Taking $l$=-4, and 1.5, we describe the effective potentials in Figs. \ref{Veffc} and \ref{Veffd}, respectively. Although the angular momenta have different signs, their patterns are quite similar. Even varying the energy, we see that there is no extremal point, which strongly indicates that TCOs do not exist for these cases. When $l$=1.6827 shown in Fig. \ref{Veffe}, the effective potential shows a similar behavior as that of Fig. \ref{Veffb}. The MSCO is present for the certain values of $a$, $l$, and $E$. For $l$=1.7, two TCOs marked with the black dots are shown in Fig. \ref{Vefff}. These disclose the characteristic behaviors of the effective potential. In summary, by varying the angular momentum, we find that there can be two TCOs, one MSCO, or no TCO.

\subsection{Vector and asymptotic behaviors}
\label{avaab}

As shown above, the TCOs are related with the zero points of the vector. Here we would like to examine the asymptotic behavior of the vector in the background of the Kerr black hole and to see whether it is consistence with the generic case shown in Sec. \ref{Circular}.

The vector can be obtained via (\ref{vectc}), which is in a complicated form and we will not show them. Expanding the vector near $\theta=0$, we have
\begin{eqnarray}
 \phi^{r}(\theta\rightarrow0)&=&-\frac{l \left(a^2 (M+r)+r^2 (r-3
   M)\right)}{\left(a^2+r^2\right)^{\frac{5}{2}}}\theta^{-1}
   +\frac{2 a \sqrt{{\Delta}} l M
   \left(a^2-3r^2\right)}{\left(a^2+r^2\right)^{\frac{7}{2}}}+\mathcal{O}(\theta),\\
 \phi^{\theta}(\theta\rightarrow0)&=&-\frac{l\sqrt{\Delta}}{\left(a^2+r^2\right)^{\frac{3}{2}}}\theta^{-2}
 \nonumber\\
 &&+\frac{\sqrt{\Delta}\left(3 a^6+a^4 \left(9 r^2-2 l^2\right)+a^2 r
   \left(9 r^3-l^2 (12 M+r)\right)+r^4
   \left(l^2+3 r^2\right)\right)}{6 l \left(a^2+r^2\right)^{\frac{7}{2}}}+\mathcal{O}(\theta),
\end{eqnarray}
and near $\theta=\pi$, we have
\begin{eqnarray}
 \phi^{r}(\theta\rightarrow\pi)&=&\frac{l \left(a^2 (M+r)+r^2 (r-3
   M)\right)}{\left(a^2+r^2\right)^{\frac{5}{2}}}(\theta-\pi)^{-1}
   +\frac{2 a \sqrt{{\Delta}} l M
   \left(a^2-3r^2\right)}{\left(a^2+r^2\right)^{\frac{7}{2}}}+\mathcal{O}(\theta-\pi),\\
 \phi^{\theta}(\theta\rightarrow\pi)&=&\frac{l\sqrt{\Delta}}{\left(a^2+r^2\right)^{\frac{3}{2}}}(\theta-\pi)^{-2}
 \nonumber\\
&&-\frac{\sqrt{\Delta}\left(3 a^6+a^4 \left(9 r^2-2 l^2\right)+a^2 r
   \left(9 r^3-l^2 (12 M+r)\right)+r^4
   \left(l^2+3 r^2\right)\right)}{6 l \left(a^2+r^2\right)^{\frac{7}{2}}}+\mathcal{O}(\theta-\pi).
\end{eqnarray}
Therefore, at $\theta=0$ and $\theta=\pi$, the direction $\arg\phi=\arctan(\phi^{\theta}/\phi^{r})$ of the vector is
\begin{eqnarray}
 \arg\phi(\theta\rightarrow0)&\sim&
 \arctan\left(-\frac{l\sqrt{\Delta}}{\left(a^2+r^2\right)^{\frac{3}{2}}}\theta^{-2}\right)\sim\arctan\left(-\theta^{-2}\right)=-\frac{\pi}{2},\\
 \arg\phi(\theta\rightarrow\pi)&\sim&
 \arctan\left(\frac{l\sqrt{\Delta}}{\left(a^2+r^2\right)^{\frac{3}{2}}}\theta^{-2}\right)\sim\arctan\left(\theta^{-2}\right)=\frac{\pi}{2}.
\end{eqnarray}
As a result, the direction of the vector is up at $\theta=\pi$ and is down at $\theta=0$. On the equatorial plane, we can also expand the vector at large $r$, which gives
\begin{eqnarray}
 \phi^{r}(\theta=\frac{\pi}{2}, r\rightarrow\infty)=\frac{M}{r^{2}}-\frac{l^2}{r^{3}}+\mathcal{O}\left(\frac{1}{r^{4}}\right).
\end{eqnarray}
Near the horizon, we denote the small positive parameter $\epsilon=r-r_{h}$. Then expanding the vector near the horizon, we obtain
\begin{eqnarray}
 \phi^{r}(\theta=\frac{\pi}{2}, \epsilon\rightarrow 0)=\frac{\sqrt{(M^2-a^2)(l^2+4M^2)}
   \left(M-\sqrt{M^2-a^2}\right)}{4 a^2 M^2}
   +\mathcal{O}(\epsilon^{\frac{1}{2}}).
\end{eqnarray}
Note that the zero order is positive. Since $\phi^{\theta}(\theta=\frac{\pi}{2})=0$, the direction of the vector is right both at the horizon and large $r$. In particular, taking $a=0$, we obtain
\begin{eqnarray}
 \phi^{r}(\theta=\frac{\pi}{2}, r\rightarrow r_{+})=\frac{1}{4 M}\sqrt{\frac{l^2}{4 M^2}+1}.
\end{eqnarray}
For the Schwarzschild black hole, one has $\kappa=1/4M$ and $g_{\varphi\varphi}^{H}=4 M^2$, so this exactly produces the result given in (\ref{kaidao}).

In summary, we conclude that the asymptotic behaviors of the vector for the Kerr black hole are the same as that of the generic discussion given above. And based on this result, we must have the fact that total topological number of the TCOs in the Kerr black hole background is zero.

\subsection{Topological charge and stability}

Here we would like to calculate the topological charge for the TCOs in the Kerr black hole background. In order to show the characteristic behavior of the unit vector $n$, we take the angular momentum $l$=1.5, 1.6827, and 1.9, respectively. For the negative value of $l$, the behaviors are quite similar. For simplicity, we take $a=0.98$ and $M=1$.

\begin{figure}
\includegraphics[width=6cm]{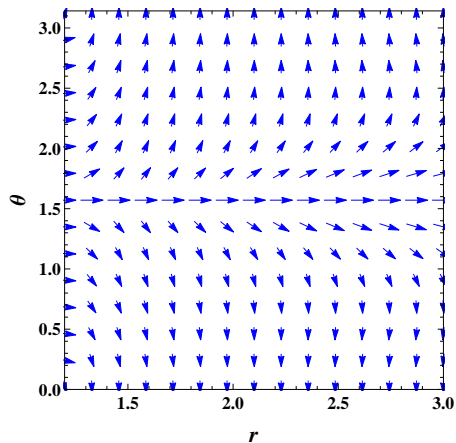}
\caption{The blue arrows represent the direction of the unit vector $n$ for $M$=1, $a=0.98$, and $l=1.5$.}\label{KerrTop}
\end{figure}

For $l$=1.5, we exhibit the unit vector $n$ in Fig. \ref{KerrTop}, where the arrows denote the direction of the vector. It is clear that for different $r$ and $\theta$, the direction of the vector is different. In the equatorial plane, one can see that the vector is towards to right, and near $\theta$=0 and $\pi/2$, the vector is up and down as expected. More importantly, there is no zero point for the vector. Thus, the topological number defining as the sum of the winding number for all the zero points is zero, i.e., $W=0$.

%%%%%%%%%%%%

\begin{figure}
\center{\subfigure[]{\label{KerrISCO}
\includegraphics[width=6cm]{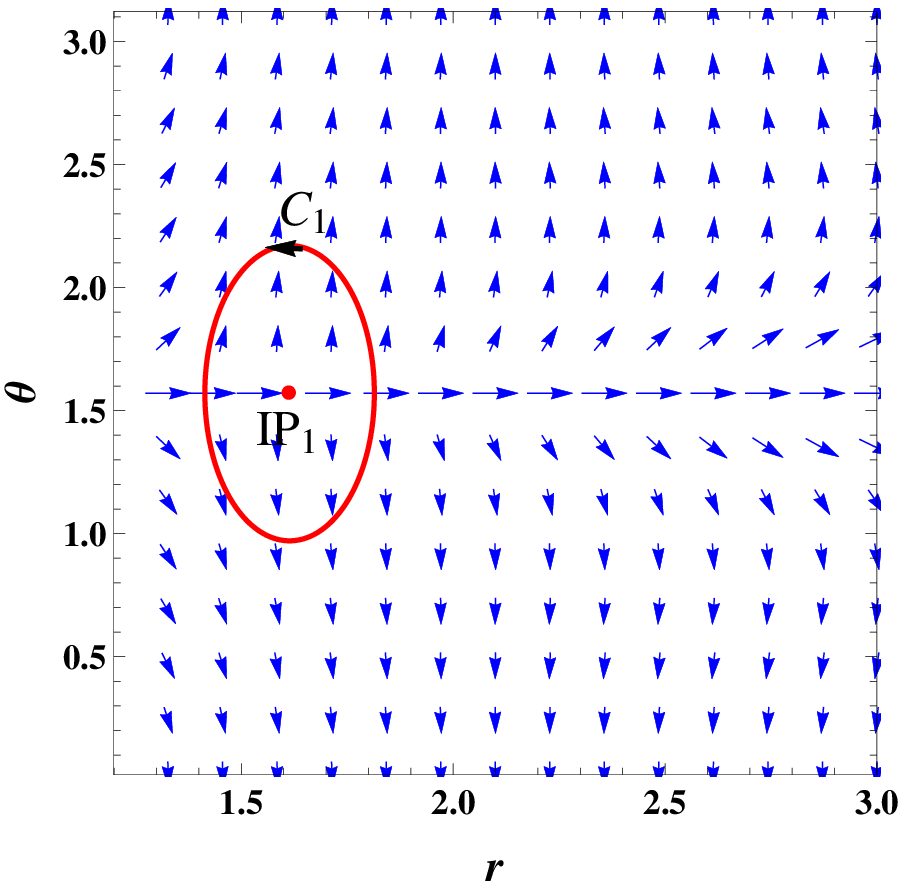}}
\subfigure[]{\label{Kerrtoch}
\includegraphics[width=6cm]{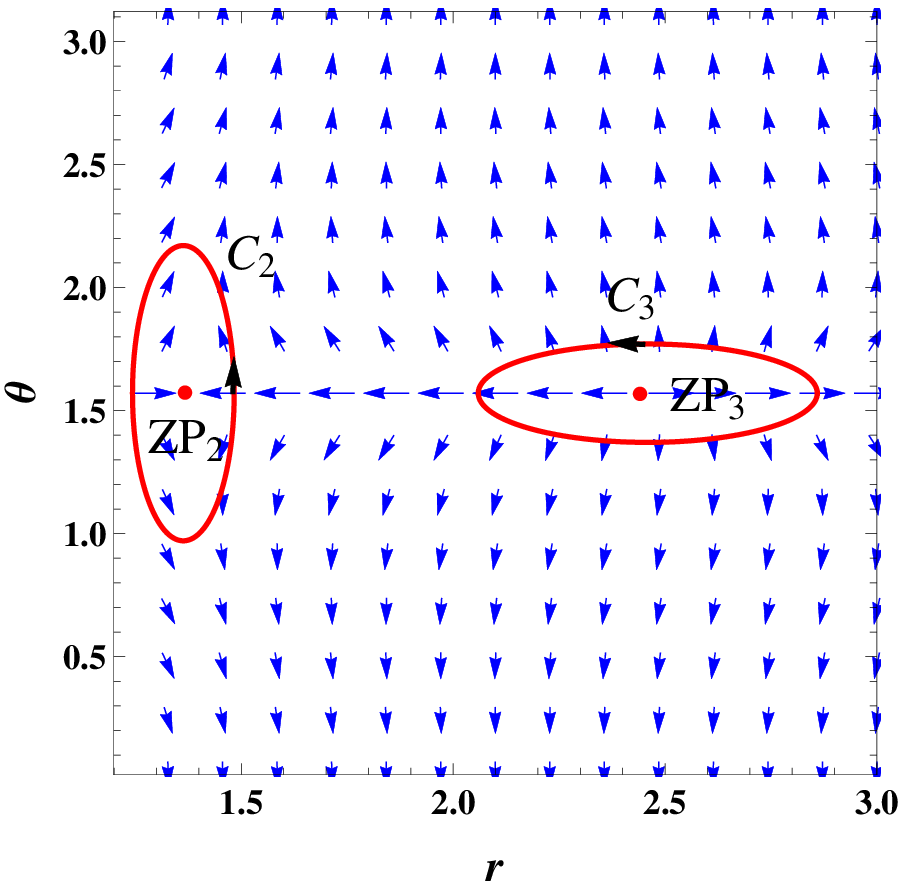}}\\
\subfigure[]{\label{KerrUTC}
\includegraphics[width=6cm]{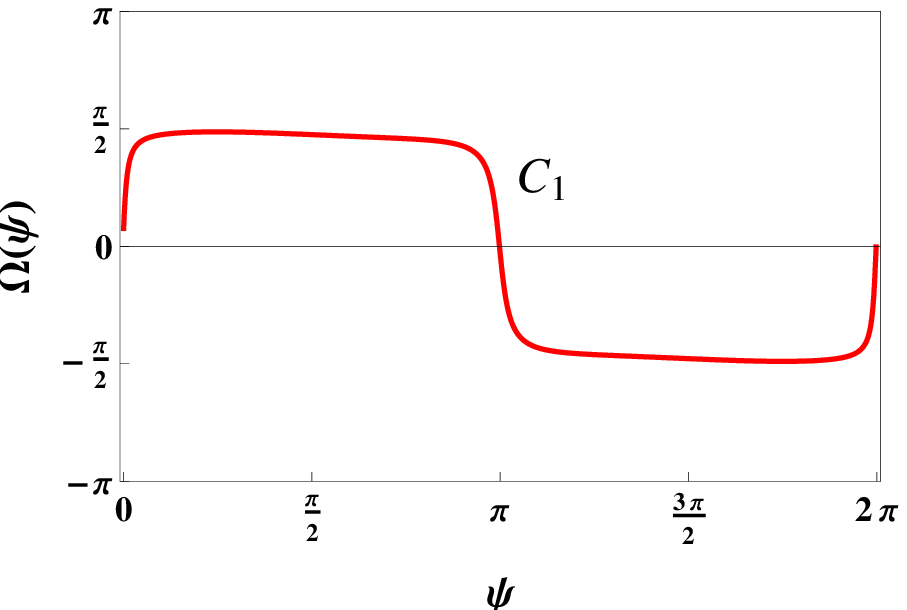}}
\subfigure[]{\label{KerrUTCtop}
\includegraphics[width=6cm]{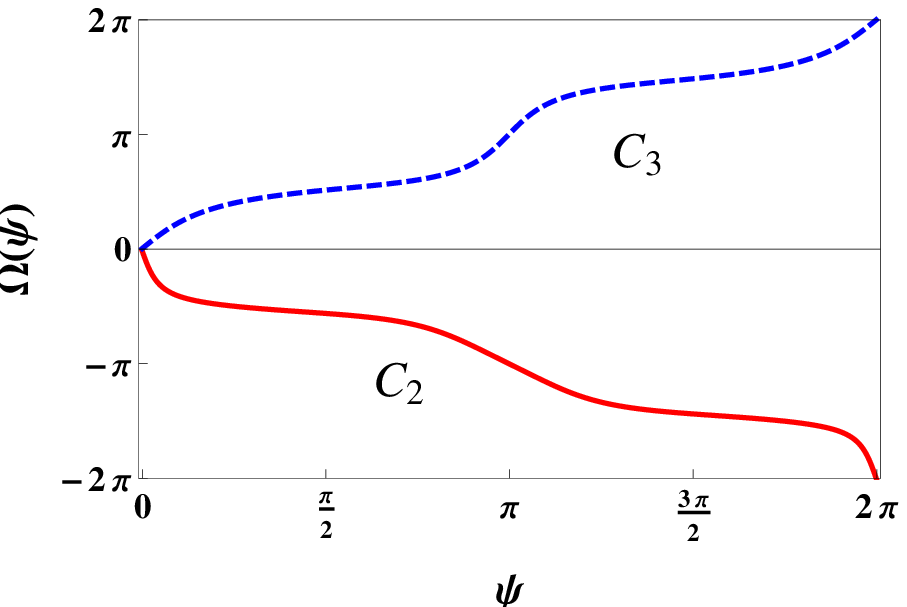}}}
\caption{(a) The blue arrows represent the direction of the unit vector $n$ for $M$=1, $a=0.98$, and $l=1.6827$. (b) The red arrows represent the direction of unit vector $n$ for $M$=1, $a=0.98$, and $l=1.9$. (c) The deflection angle $\Omega(\psi)$ along the closed loop $C_{1}$. (d) The deflection angle $\Omega(\psi)$ along the closed loops $C_{2}$ (red solid curve) and $C_{3}$ (blue dashed curve). ``IP" and ``ZP" denote the locations of the ISCO and the zero points of the vector.}\label{ppKerrtoch}
\end{figure}

If we set $l=1.6827$, the MSCO or ISCO will be present. For this particular case, we show the unit vector $n$ in Fig. \ref{KerrISCO}. Its behavior is similar to that of Fig. \ref{KerrTop}. However, one IP$_1$ point can be discovered at $r=1.61$, which actually relates to the MSCO of the black hole. To calculate the winding number for the MSCO, we construct a closed loop parametrized by the following parametrized form
\begin{eqnarray}
\left\{
\begin{aligned}
 r&=c_1\cos\psi+c_0, \\
 \theta&=c_2\sin\psi+\frac{\pi}{2}.
\end{aligned}
\right.\label{pfs}
\end{eqnarray}
For the closed loop $C_1$, we have $(c_0, c_1, c_2)$=(1.61, 0.2, 0.6). Along the loop $C_1$, we calculate the deflection angle $\Omega(\psi)$ defined as \cite{Wei2019}
\begin{equation}
 \Omega(\psi)=\int_{C_{1}} \epsilon_{ab}n^a\partial_{i}n^bdx^{i}.
\end{equation}
The result is listed in Fig. \ref{KerrUTC}. With the increase of $\psi$ from 0 to $2\pi$, $\Omega(\psi)$ first increases, then decreases, and finally increases. However, it does not make one complete loop. So the winding number for the MSCO is zero, and thus $W=0$.

As a third characteristic example, we take the angular momentum $l=1.9$. The unit vector $n$ is exhibited in Fig. \ref{Kerrtoch}. Quite different from previous cases, there are two zero points ZP$_2$ and ZP$_{3}$ located at $r=$1.36 and 2.46. In order to calculate their winding numbers, we construct two closed loops C$_2$ and C$_3$ with parametrized coefficients $(c_0, c_1, c_2)$=(1.36, 0.12, 0.6) and (2.46, 0.4, 0.2), respectively. Along these loops, we show the deflection angle in Fig. \ref{KerrUTCtop}. With the increase of $\psi$, $\Omega(\psi)$ decreases for C$_2$ and increases for C$_3$. When $\psi$ vary from 0 to $2\pi$ denoting a complete loop, we find that the vector makes one loop clockwise along C$_2$, and counterclockwise along C$_3$. Therefore, we obtain $w_{ZP_2}$=-1 and $w_{ZP_3}$=1 for the zero points ZP$_2$ and ZP$_{3}$. Considering that the TCOs with small and large radii, respectively, are unstable and stable, one can conclude that the unstable TCOs have negative winding number, while the stable ones have positive winding number. Since the topological number is the sum of the winding numbers of all the zero points, we still have $W=-1+1=0$. This indicates that the total topological number still holds unchanged for this case even when two zero points are present.

\subsection{Evolution of control parameter}

In the above, we have shown several characteristic behaviors of the vector for different values of the angular momentum acting as a control parameter. Here we would like to consider the TCOs and their topological number under the evolution of the control parameter.

\begin{figure}
\center{\subfigure[]{\label{CPara}
\includegraphics[width=6cm]{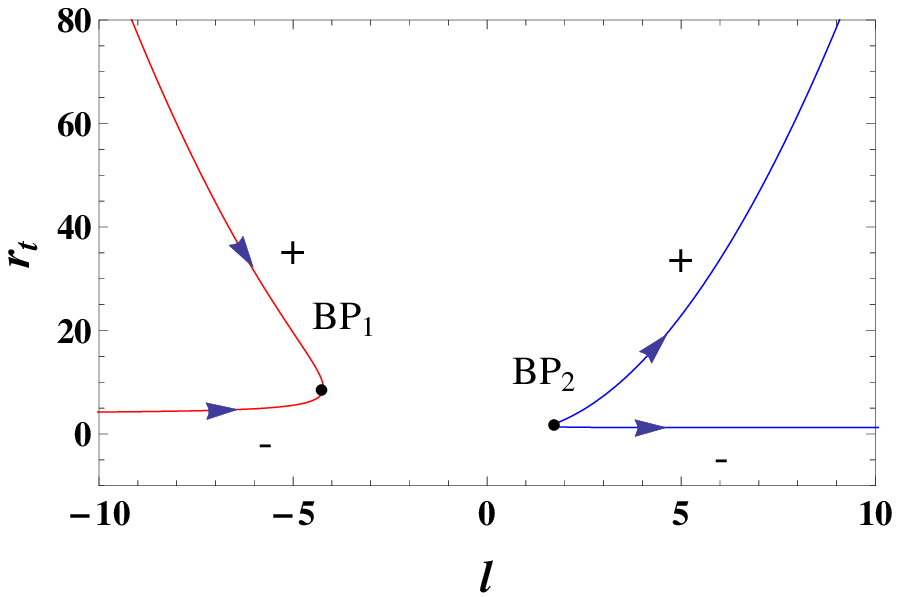}}
\subfigure[]{\label{CParaTOP}
\includegraphics[width=6cm]{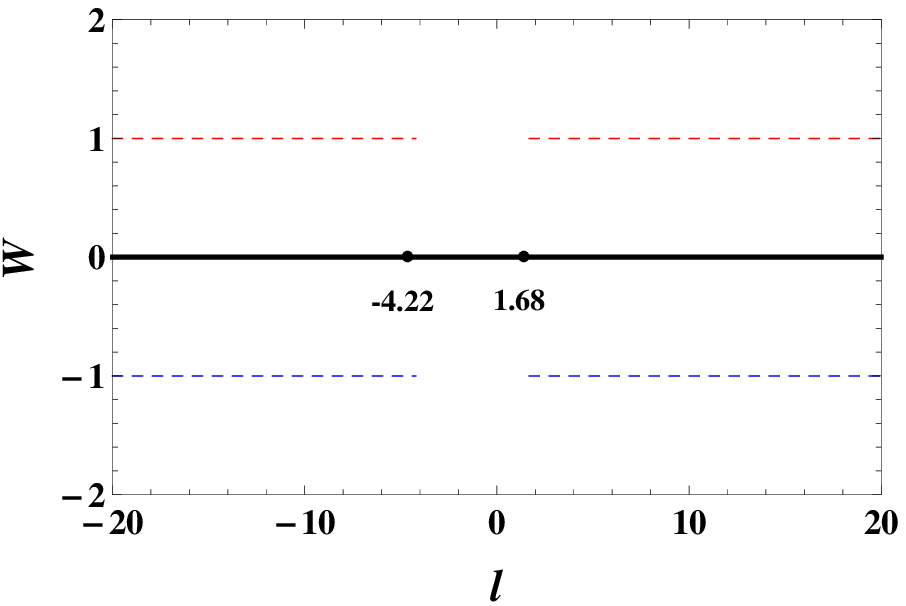}}}
\caption{(a) The evolution of the radius $r_{\text t}$ of the TCO as the angular momentum $l$. BP$_1$ and BP$_2$ are two bifurcation points. The signs ``+" and ``-" denote that the branches have positive and negative winding numbers, respectively. (b) The total topological number $W$ (black solid line) relating to the TCOs as a function of the angular momentum $l$. Dashed lines denote the winding numbers of these TCO branches. These black dots are for the bifurcation points (BP). The black hole parameters are set to $M$=1 and $a=0.98$.}\label{ppCParaTOP}
\end{figure}

We plot the radii of the TCOs as a function of the angular momentum $l$ in Fig. \ref{CPara}, which can vary from $-\infty$ to $+\infty$. There are four different branches. The signs ``$+$" and ``$-$" denote that the branches have positive and negative winding numbers, respectively. For sufficiently negative angular momentum, there are two branches of TCOs. The upper branch has a positive winding number, while the lower one has a negative winding number; their sum vanishes. With the increase of the angular momentum, these two branches approach each other, and coincide at BP$_1$ at $l$=-4.22. In the region $l\in(-4.22, 1.68)$, no TCO branch can be observed, and thus the topological number is zero. Further increasing the angular momentum, we observe that two branches extend from the point BP$_2$ at $l$=1.68. The upper and the lower ones have positive and negative winding numbers, respectively. The sum of the winding number is still zero.

From this behavior of $r_{\text t}$ of the TCOs, we can find that the topological number always stays at zero for different values of the angular momentum $l$. We show the details in Fig. \ref{CParaTOP}. The dashed lines denote the winding number of different TCO branches, and the black solid line is the total topological number of the TCOs. It implies that the total topological number always vanishes.

Moreover, from Fig. \ref{CPara}, one can clearly see that the points BP$_1$ and BP$_2$ are two bifurcation points. Actually, they are the MSCOs for the Kerr black hole with $a$=0.98 and $M$=1. Near these two points, we have
\begin{equation}
 \frac{d^2l}{dr_{\text t}^{2}}=-0.05\quad \text{and}\quad 2.31,
\end{equation}
which indicates that BP$_1$ and BP$_2$ are the annihilated and generated points, respectively. This is also consistent with the result shown in Fig. \ref{CPara}.

Before ending this section, we briefly summarize our results. For the Kerr black hole, the topological number relating to the TCOs vanishes, indicating if the TCOs exist, they always come in pairs for the fixed angular momentum. Note that although we here take $a$=0.98 and $M$=1 for an example, this result holds for other values of the black hole mass and spin.

\section{Control parameter and topology}
\label{Control}

As shown above, we take the angular momentum $l$ as the control parameter to construct the topology. Since the energy $E$ is also an alternative parameter for the massive test particle, it is interesting to consider this case.

Here we need to reformulate the effective potential $\mathcal{V}$ as
\begin{eqnarray}
 \mathcal{V}=-\frac{g_{tt}}{B}(l-l_{1})(l-l_{2}),\label{ppotent2}
\end{eqnarray}
where $l_{1,2}$ are given by
\begin{eqnarray}
 l_{1}&=&\frac{-E g_{t\phi}+\sqrt{B}\sqrt{g_{tt}+E^2}}{g_{tt}},\\
 l_{2}&=&-\frac{E g_{t\phi}+\sqrt{B}\sqrt{g_{tt}+E^2}}{g_{tt}}.
\end{eqnarray}
It is obvious that $l_1\leq0$ corresponds to the retrograde orbits. Otherwise, $l_2$ corresponds to prograde orbits. So the conditions of the TCO reduces to
\begin{eqnarray}
 l=l_1,\quad \text{and}\quad \partial_{r}l_1=0,
\end{eqnarray}
for retrograde cases, and
\begin{eqnarray}
 l=l_2,\quad \text{and}\quad \partial_{r}l_2=0,
\end{eqnarray}
for prograde cases. As a result, we need to construct two different vectors to describe the TCOs for each rotating sense.\\

\noindent Case I: the vector $\phi_1$ is constructed as
\begin{eqnarray}
 \phi_{1}^{r}=\frac{\partial_{r} l_{1}}{\sqrt{g_{rr}}},\quad
 \phi_{1}^{\theta}=\frac{\partial_{\theta} l_{1}}{\sqrt{g_{\theta\theta}}}.\label{vectca}
\end{eqnarray}
Via solving the zero point of $\phi_1$, we obtain that the corresponding energy is $E=E_{-}$ given in (\ref{eee}). Plugging $E_{-}$ into $l_1$, one can obtain the angular momentum $l=l_{-}$ as expected.\\

\noindent Case II: the vector $\phi_2$ reads
\begin{eqnarray}
 \phi_{2}^{r}=\frac{\partial_{r} l_{2}}{\sqrt{g_{rr}}},\quad
 \phi_{2}^{\theta}=\frac{\partial_{\theta} l_{2}}{\sqrt{g_{\theta\theta}}}.\label{vectcb}
\end{eqnarray}
For such case, we can obtain the zero point of $\phi_2$ at $E=E_{+}$ and $l=l_{+}$.

\begin{figure}
\center{\subfigure[]{\label{CPare}
\includegraphics[width=6cm]{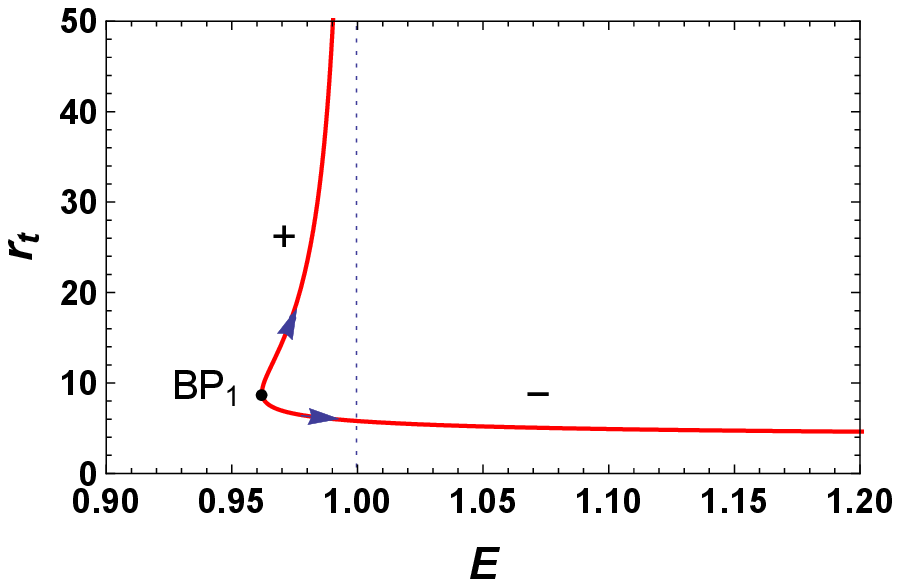}}
\subfigure[]{\label{CParettb}
\includegraphics[width=6cm]{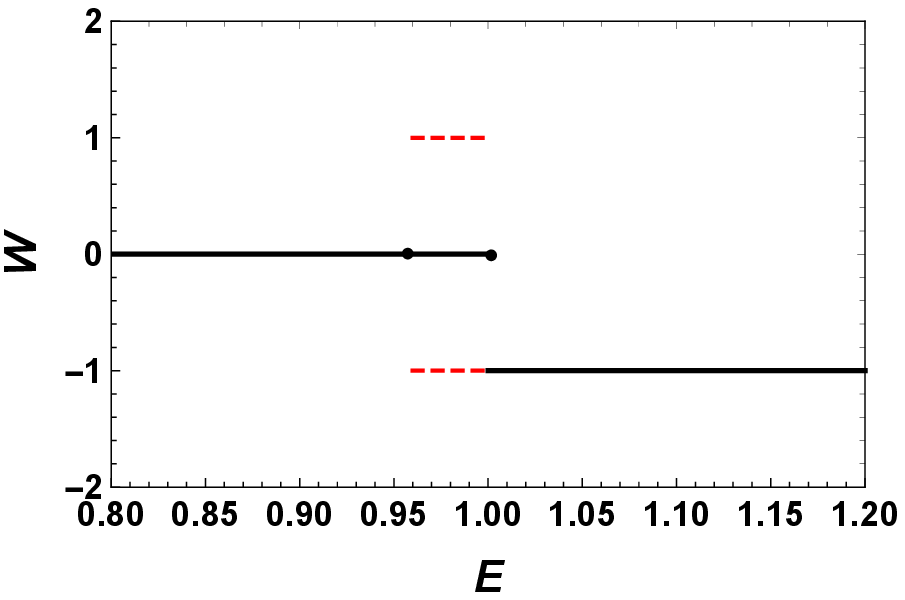}}}
\caption{Topological behaviors for the retrograde TCOs. (a) The evolution of the radius $r_{\text t}$ of the TCO as the control parameter $E$. BP$_1$ is a bifurcation point at $E=0.96$. The signs ``+" and ``-" denote that the branches have positive and negative winding numbers, respectively. (b) The total topological number $W$ (black solid line) relating to the TCOs as a function of the energy $E$. Dashed lines denote the winding numbers of these TCO branches. The first and second black dots are for the bifurcation point and phase transition point. The black hole parameters are set to $M$=1 and $a=0.98$.}\label{ppCPare}
\end{figure}

\begin{figure}
\center{
\subfigure[]{\label{CParec}
\includegraphics[width=6cm]{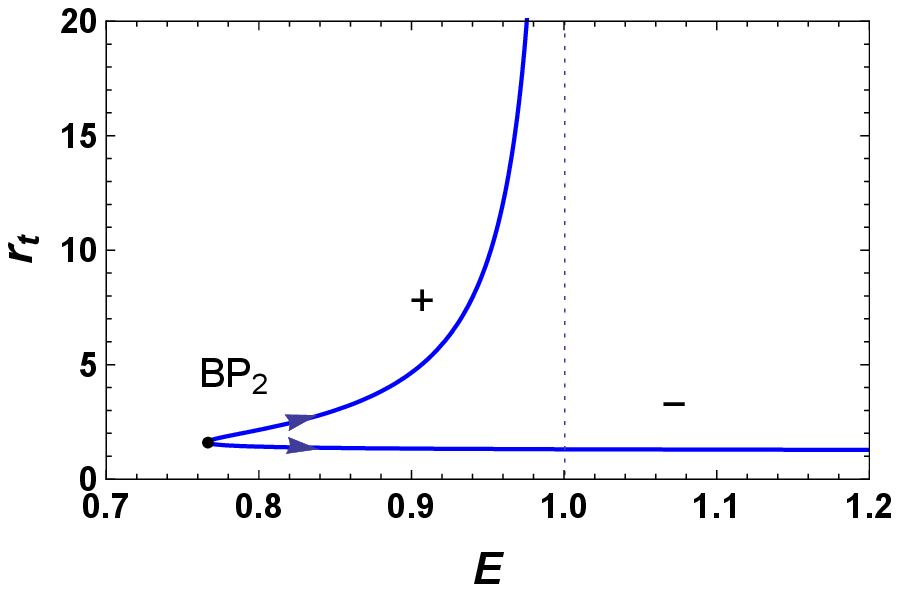}}
\subfigure[]{\label{CParetaadp}
\includegraphics[width=6cm]{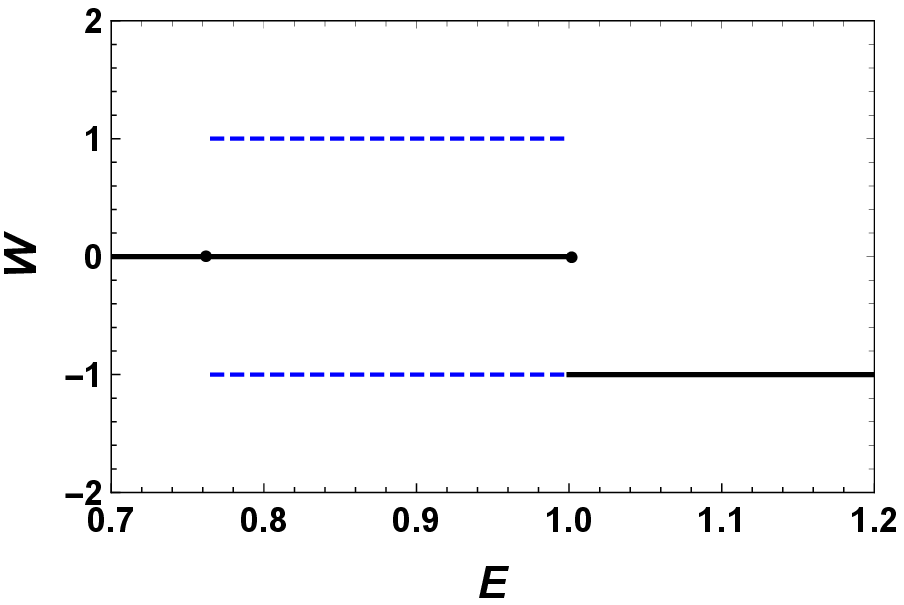}}}
\caption{Topological behaviors for the prograde TCOs. (a) The evolution of the radius $r_{\text t}$ of the TCO as the control parameter $E$. BP$_2$ is a bifurcation point at $E=0.77$. The signs ``+" and ``-" denote that the branches have positive and negative winding numbers, respectively. (b) The total topological number $W$ (black solid line) relating to the TCOs as a function of the energy $E$. Dashed lines denote the winding numbers of these TCO branches. The first and second black dots are for the bifurcation point and phase transition point. The black hole parameters are set to $M$=1 and $a=0.98$.}\label{CParetaad}
\end{figure}

For each rotating sense, we can follow the treatment shown in Sec. \ref{Topology} to obtain the topological number by analyzing the asymptotic behaviors of the vectors $\phi_1$ and $\phi_2$. Owing to that the process is quite similar, we just present our result here. For both the retrograde and prograde TCOs, we observe that, the topological number $W$ vanishes for $E\in(0, 1)$, while when the energy is larger than one, the topological number is $W$=-1. This is mainly because that the direction of vectors $\phi_{1,2}^{r}$ at large $r$ changes when the energy crosses $E=1$. So there is a topological phase transition at $E=1$ for both the rotating senses. As we know, in an asymptotically-flat spacetime, the motion of the massive particles with energy $E<1$ is bounded and cannot reach the infinity. Instead, these particles with $E>1$ still keep moving at the infinity, and bounded orbits cannot be observed. So the critical case is that these particles with energy $E=1$ can stay at rest in the infinity. As a result, the energy $E$=1 plays a characteristic point to distinguish these particles which cannot reach or freely move at infinity. It also measures whether the bounded orbits exist or not. From our topological approach, we clearly see that this physical feature can be well-described by the topological phase transition at $E=1$.

In order to clearly show them, we exhibit our results in Figs. \ref{ppCPare} and \ref{CParetaad} for the Kerr black hole with $M$=1 and $a$=0.98 for these two rotating senses. In Figs. \ref{CPare} and \ref{CParec}, we plot the evolutions of the radii $r_{\text t}$ of the TCOs as the control parameter $E$ for the retrograde and prograde cases, respectively. For the retrograde case, when $E<0.96$, there is no TCO branch, and thus the total topological number is zero. When $E=0.96$, two branches of negative and positive winding numbers emerge from the point BP$_1$, and for the prograde case, we observe a similar behavior near BP$_2$ at $E=0.77$.

 Near these two points BP$_1$ and BP$_2$, we have
\begin{equation}
 \frac{d^2E}{dr_{\text t}^{2}}=0.0018\quad \text{and}\quad 0.7631,
\end{equation}
which indicates that all these bifurcation points are generated points. Note that the bifurcation points BP$_1$ and BP$_2$ are exactly the ones shown in Fig. \ref{CPara}.

As the energy approaches one, these two branches possessing positive winding number tend to infinity $r_{\text t}\rightarrow\infty$. Nevertheless, the total topological number always stays at zero as expected. When the energy is larger than one, only one branch is left for each rotating sense, but both them have negative winding number $w=-1$. Therefore, the topological number is $W=-1$ for each case. These behaviors suggest that there is a topological phase transition at $E=1$, where the topological number $W$ turns from 0 to -1. Note that the values of the winding number correspond the stability of the TCOs. As we have shown, the TCO with a negative winding number are unstable. Any tiny perturbation will make the particle deviate from the TCO, so for the particles with $E>1$, only unstable TCOs exist. Although these TCOs are very limited for the study of astronomical observable effects, it can help us understand the complete structure of TCOs in the black hole backgrounds.

We show the topological number $W$ in Figs. \ref{CParettb} and \ref{CParetaadp} for the retrograde and prograde cases, respectively. It is clear that the topological number stays at zero when $E<1$ for both cases, regardless of the bifurcation point. When $E>1$, the topological number turns to -1 as expected. Thus, there is a topological phase transition at $E$=1.

Noting that in our discussion above, we only take $a=0.98$. It is worth examining whether topological pattern and phase transition depend on the mass and spin of the black hole. As shown in Sec. \ref{avaab}, even though the black hole mass and spin shall modify the coefficients of the expansion of the vector, they do not affect the asymptotic behaviors. Thus, the total topological number is independent of the black hole mass and spin. Nevertheless, the locations of the TCOs and the MSCOs will be shifted. On the other hand, the topological phase transition at $E$=1, is only dependent of the behavior of $\phi^r$ at the infinity, so such a phase transition holds for arbitrary values of the black hole mass and spin.

\section{Conclusions and discussions}
\label{Conclusion}

In this paper, we constructed a well-behaved topology for the TCOs in a generic black hole spacetime, even though the TCOs are closely dependent of the energy and angular momentum of the massive particle.

At first, we treated the angular momentum as the control parameter. Similar to the treatment of the photon \cite{Cunhaa,Cunhab}, we started with the timelike geodesics and constructed a vector $\phi$ by using the effective potential. The TCO exactly locates at the zero point of $\phi$. Thus, one can endow each of them with a local topological charge-the winding number, for the TCO. Moreover, we studied the asymptotic behaviors of $\phi$ near the boundaries in the $r$-$\theta$ plane. Via examining the axis limit, horizon limit, and asymptotic limit, we found the value of the angular momentum does not change the asymptotic behaviors of the vector. This strongly implies that the topological approach can be applied to the TCOs, just like that for the LRs.

Then we considered the global and local topologies relating with the TCOs. From the local viewpoint, we found that the positive or negative winding number, respectively, corresponds to stable or unstable TCO. In particular, the MSCO is just the bifurcation point that the stable and unstable TCOs can generate from or annihilate into. Nevertheless, the sum of the winding number stays the same before or after the bifurcation point. From the global viewpoint, we observed that the total topological number, defined as the sum of the winding numbers of all the zero points always vanishes for a generic black hole spacetime. It suggests that the TCOs with the fixed angular momentum come in pairs with one being stable and another being unstable. It is worth pointing out that this result is universal and is independent of the angular momentum of the particle and the black hole parameters. However, the total topological number for the LR is -1, indicating there exists a standard LR at least for a generic black hole.

As a specific example, we studied the Kerr black hole. As expected, the vector is outwards at $\theta$=0 and $\pi$. Also the vector is to the right both at $r=r_{\text{h}}$ and $\infty$. This pattern indicates that the total topological number of the TCO is $W$=0 for the Kerr black hole. So, if one treats the angular momentum as the control parameter, the stable and unstable TCOs always come in pairs. Of particular interest is that the two MSCOs for both the rotating senses act as the bifurcation points. The one with negative angular momentum is an annihilated point and the other one with positive angular momentum is a generated point. Although for different values of the angular momentum, there may two or zero TCO branches, the total topological number is always zero as expected.

Besides the angular momentum, the energy of the test particle is also an alternative control parameter. Hence, we examine the corresponding topology for both the rotating senses, respectively. The results show that the total topological number is $W$=0 for $0<E<1$, while $W$=-1 for $E>1$. This suggests that there exists a topological phase transition at $E$=1 both for the retrograde and prograde cases, which actually measures different motions of the massive particles at infinity. Moreover, the MSCOs in this case exactly correspond to two generated points.

Before ending this paper, we would like to address three points for the topology of the TCO that are different from the LR: i) For the LR, the energy and angular momentum are linear, while for the TCO, they have a nonlinear relation; ii) When the black hole background is fixed, the radius of the LR is independent of the parameters of the photon, while for the TCO, its radius closely depends on the test particle. It appears that the degeneracy of the circular orbit is broken when the test particle gets mass, see Fig. \ref{ppSkf} for an example; iii) The total topological number is $W$=-1 for LR, while 0 for TCO. This implies that for a generic black hole, there exists at least one standard LR, while the TCOs always come in pairs for the fixed angular momentum. Here we developed a general topological treatment for the TCOs, and applied it to the Kerr black hole. We believe that more information about the TCOs could be disclosed for other rotating and nonrotating black holes in GR or modified theories of gravity by employing this topological approach, and more general black hole solutions without $\mathcal{Z}_2$ will be further investigated.

\section*{APPENDIX: REMOVAL OF DEGENERACY OF THE CIRCULAR ORBIT}

\begin{figure}
\center{
\includegraphics[width=14cm]{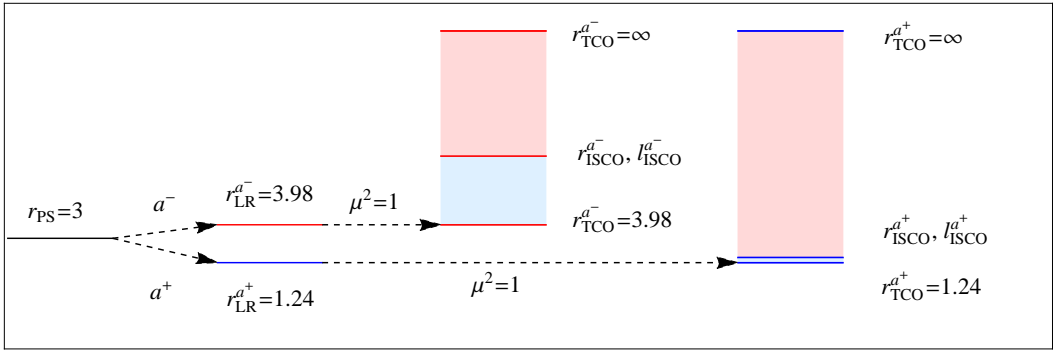}}
\caption{Schematic illustration of the removal of degeneracy of the circular orbit with $M$=1 and $a$=0.98. The PS of the Schwarzschild black hole is broken by the black hole spin $a$ and particle mass $\mu^2$. Light red and blue color regions are for the stable and unstable TCOs. ($r_{ISCO}^{a^{\pm}}$, $l_{ISCO}^{a^{\pm}}$)=(1.61, 1.68) and (8.94, -4.22).}\label{ppSkf}
\end{figure}

Taking $M$=1 and $a=0.98$ as an example, we shown the removal of degeneracy of the circular orbit due to the black hole spin and mass of the test particle in Fig. \ref{ppSkf}. For the Schwarzschild black hole, the radius of the photon sphere $r_{PS}$=3. When the black hole spin is present it is split into two; the prograde and retrograde LRs located at $r_{LR}$=1.24 and 3.98, respectively. Here $a^{\pm}$ denotes that the black hole spin $a$ and angular momentum $l$ have the same or opposite direction. When the test particle gains mass, its motion will be described by the timelike geodesics rather than the null geodesics. Each LR will turn into a continuous set of TCOs with the radius extending from the value of LR to infinity. The unstable and stable TCOs are separated by the MSCOs or ISCOs. Note that the radii of the unstable TCOs are bounded by the ISCOs and LRs. For the $a^+$ branch, the particle angular momentum starts at the MSCO with $l$=1.68 to $r_{TCO}^{a^{\pm}}$=1.24 and $\infty$ with $l=\infty$, and for the $a^-$ branch, the particle angular momentum starts at the MSCO with $l$=-4.22 to $r_{TCO}^{a^{\pm}}$=3.98 and $\infty$ with $l=-\infty$.

\section*{Acknowledgements}
The authors would like to thank Dr. Jian-Zhi Yang for the useful discussions. This work was supported by the National Natural Science Foundation of China (Grants No. 12075103, No. 11875151 and No. 12247101).

\end{document}